\documentclass[aps,twocolumn,groupedaddress,nofootinbib] {revtex4-1}
\usepackage{amsmath}
\usepackage{amsfonts}
\usepackage{amssymb}
\usepackage{graphicx}
\usepackage{caption}
\usepackage{subcaption}
\usepackage{float}
\usepackage{braket}
\allowdisplaybreaks
\newcommand {\bp}{\begin{pmatrix}}
\newcommand {\ep}{\end{pmatrix}}
\newcommand{\be}{\begin{equation}} \newcommand{\ee}{\end{equation}}
\newcommand{\bea}{\begin{eqnarray}}\newcommand{\eea}{\end{eqnarray}}

\begin{document}

\title{Quantum Integrability and Chaos in periodic Toda Lattice with Balanced Loss-Gain}
\author{Supriyo Ghosh}
\email[]{supriyoghosh711@gmail.com}
\author{Pijush K. Ghosh}
\email[Corresponding Author: ]{pijushkanti.ghosh@visva-bharati.ac.in}
\affiliation{ Department of Physics, Siksha-Bhavana, Visva-Bharati, 
Santiniketan, PIN 731 235, India.}
\date{\today}

\begin{abstract}
We consider equal-mass quantum Toda lattice with balanced loss-gain for two
and three particles. The two-particle Toda lattice is integrable and two integrals of
motion which are in involution have been found. The bound-state energy and the corresponding
eigenfunctions have been obtained numerically for a few low-lying states. 
The three-particle quantum Toda lattice with balanced loss-gain and velocity mediated coupling
admits mixed phases of integrability and chaos depending
on the value of the loss-gain parameter. We have obtained analytic expressions for two integrals
of motion which are in involution. Although an analytic expression for the third integral has not
been found, the numerical investigation suggests integrability below a critical value of
the loss-gain strength and chaos above this critical value.
The level spacing distribution changes from the Wigner-Dyson to the Poisson distribution
as the loss-gain parameter passes through this critical value and approaches zero. An identical
behaviour is seen in terms of the gap-ratio distribution of the energy levels. The existence of
mixed phases of quantum integrability and chaos in the specified ranges of the loss-gain parameter
has also been confirmed independently via the study of level repulsion and complexity in higher
order excited states.
\end{abstract}
\maketitle
\tableofcontents{}
\section{Introduction}

The Toda lattice, a system of particles interacting via exponential potentials, has long been
the subject of extensive research\cite{toda1,toda2, toda3}. Its remarkable properties, such as
solitonic solutions\cite{date1976} and the integrability of its equations of motion\cite{flaschka,
henon,agrotis}, have made it a pivotal topic in the study of integrable systems and mathematical
physics. Toda lattice is important in the context of mathematical modelling of many physical phenomena
like heat propagation in lattice system\cite{toda4,sataric}, dynamics of DNA\cite{muto}, peptide
bonds in the $\alpha$-helix\cite{dovivio}, laser dynamics\cite{oppo,lien,cialdi}. Various generalizations 
and modification of Toda lattice  having realistic physical applications and mathematical importance 
have been considered\cite{habib,casati,vergara,ezawa,proy}. Truncated Toda potential perturbed by weak 
friction and noise is important in galactic dynamics\cite{habib}. Chaotic behaviour is seen for 
Toda lattice with unequal masses\cite{casati}.

The Balanced Loss-Gain(BLG) system is defined as the one in which the flow preserves the volume
in the position-velocity state 
space, although the individual degrees of freedom may be subjected to gain or loss\cite{pkg}. The 
system is non-dissipative and may admit a Hamiltonian. A novel feature of such systems is the
existence of (quasi-)periodic solutions within some regions of the parameter-space, and have been
studied extensively in the context of $\mathcal{PT}$-symmetry\cite{bender, bender2, cuevas,
barashenkov,ds1,khare1,ds2,ds3,ds4,proy1,proy2, proy, sg1,sg2,pkg2}. The Hamiltonian formulation 
of generic BLG systems has been discussed in Ref. \cite{ds2, ds3, ds4, pkg1} and a few such examples
include systems with nonlinear interaction\cite{barashenkov,ds1,khare1,ds2,ds3,ds4},
many particle systems\cite{ds2,ds3,ds4}, systems with space dependent loss-gain term\cite{ds3}, systems 
with Lorentz interaction\cite{pkg1}, Hamiltonian chaos\cite{proy1,proy2,proy}, oligomers\cite{pkg} etc.
The formalism has been extended beyond the mechanical systems \textemdash nonlinear Schr\"odinger
equation\cite{pkg2, sg1, sg2} and nonlinear Dirac equation\cite{pkg} with BLG have been studied
from the viewpoint of exact solvability and existence of solutions bounded in time.

A recent addition to the growing list of generalized Toda systems is a Toda lattice with BLG\cite{proy}.
It has been shown that a two-particle Toda system with BLG is integrable and analytic expressions for
two integrals of motion which are in involution have been found. Periodic solutions of the equations
of motion have been found numerically. A three particle Hamiltonian Toda system with BLG and
Velocity Mediated Coupling(VMC) has been shown to admit mixed phases of integrability and chaos.
Two of the integrals of
motion which are in involution have been found analytically. Although the third integral of motion has
not been found analytically, which is required for showing integrability, the numerical studies
reveal integrability below a critical value of the
loss-gain parameter and chaos above this critical value. The existence of mixed phases of integrability
and chaos in the system has been established through the study of sensitivity to the initial
conditions, Poincar\'e Sections, Lyapunov exponent, power-spectra and auto-correlation functions\cite{proy}.
A non-Hamiltonian Toda system with BLG has also been shown to admit chaotic behaviour and chaos in the
system is solely induced due to BLG. 

The purpose of the article is to study the Hamiltonian Toda system with BLG and VMC 
of Ref. \cite{proy} in the quantum region. The two-particle quantum Toda system is shown
to be integrable via the construction of two integrals of motion which are in involution.
The translation invariance of the system allows to separate the center of mass motion
and the Sch\"odinger equation in the relative coordinate may be interpreted as that of
a particle moving in an inverted harmonic oscillator plus a cosine hyperbolic potential.
The effective potential is of a single-well or a symmetric/asymmetric double-well depending
on the parameters of the system. The quantum bound states have been obtained numerically.

The three particle Hamiltonian Toda lattice is translation invariant. The center of
mass motion is separated out in the Jacobi coordinate and the effective Hamiltonian may
be interpreted as that of a particle moving in a two dimensional potential,
consisting of anisotropic harmonic oscillators and the Toda Potential, and subjected
to an external uniform magnetic field with its magnitude being proportional to the loss-gain
parameter. The system is exactly solvable in the limit, in which the Toda potential reduces
to that of coupled oscillators, such that the starting Hamiltonian describes coupled oscillators
with BLG and VMC. The eigenvalues and eigenfunctions
are obtained analytically in this particular limit which are given by two decoupled anisotropic
oscillators. 
 
In general, the quantum problem is not amenable to analytic solution. 
The conserved quantity corresponding to the translation invariance may be interpreted as a
generalized momentum which commutes with the Hamiltonian, thereby,
two of the quantum integrals of motion are found analytically. The third integral of motion that
is required to establish integrability has not been found analytically so far. However,
the numerical investigations suggest integrability below a critical value of the loss-gain parameter
and chaos above this critical value, as is the case for the corresponding classical system. It may be
recalled in this context that a link between the classical and quantum chaos may be explored by studying
statistical properties of energy levels based on Random Matrix Theory(RMT)\cite{haake,bgs1984}.
In particular, for Gaussian Orthogonal Ensemble(GOE),
the level statistics of quantum Hamiltonian with integrable classical counterpart follows Poisson distribution
\cite{berry1977,montambaux,poilblanc,gaspard, georgeot,kudo, deguchi,santos,gubin}, while in the chaotic
region it follows the Wigner distribution.
The level statistics is described in terms of the spacing of nearest-neighbour eigen-energies, 
and unfolded data is used in the process instead of the raw data. There is an alternative method
to circumvent the problems associated with unfolding procedure in some cases where the statistics
of ratios of the energy-gaps for consecutive levels is studied. The statistics of gap-ratio
of quantum Hamiltonian with integrable classical counterpart follows Poisson distribution, while
it follows Wigner distribution in the chaotic region\cite{oganesyan}.

In this article, we establish the mixed phases of integrability and chaos by studying the level
spacing distribution as well as gap-ratio distribution. The quantum transition from the chaotic
to the integrable region is observed when the loss-gain strength 
crosses a critical value and goes to zero \textemdash the level spacing as well as the gap-ratio
distributions smoothly changes from the Wigner-Dyson distribution and tends to follow the Poisson
distribution. We also show the level repulsion phenomena in the energy spectra in both the two and
three particle Toda lattice. It is shown through the graphical presentations that the degree of level
repulsion is large in the case of the three particle system. The quantum transition from chaotic to
integrable region is also confirmed independently by studying the complex behaviour of
higher order excited state wave functions.

The plan of this article is the following. The two-particle Toda system is studied in the next section.
The three particle Toda system is presented in Sec. III and the exactly solvable limiting case is described
in Sec. III.A. The numerical results for the general quantum problem is described in Sec. III.B. Finally,
the results are summarized with discussions in Sec. IV.

\section{Two Particle Toda lattice}

The Hamiltonian of the Periodic Toda lattice is given as\cite{proy} 
\bea
H & = & 2 \left ( p_1 + \frac{\gamma}{2} x_2 \right) \left( p_2 - \frac{\gamma}{2} x_1 \right) \nonumber \\
  & + & \frac{a}{b} \left\{ e^{b(x_1-x_2)} + e^{b(x_2 - x_1)} - 2 \right\},
\label{ham1}
\eea
\noindent where $\gamma$ is the strength of the BLG term and $a$,$b$ are the 
strengths of nonlinear interaction. The canonical conjugate momenta to the coordinates
$(x_1,x_2)$ are defined as  $(p_1,p_2)$. The Hamiltonian (\ref{ham1}) is $\mathcal{PT}$-symmetric 
where the parity transformation $\mathcal{P} : $ \ $x_1 \leftrightarrow x_2, p_1 \leftrightarrow p_2$ and 
the time-reversal operation $\mathcal{T} :  (x_1, x_2) \rightarrow (x_1, x_2)$, $ (p_1, p_2) \rightarrow
(-p_1, - p_2)$.  In the limit of $a \rightarrow \infty, b \rightarrow 0$ such that $ab \equiv\omega^2$,
the above Hamiltonian reduces to that of coupled oscillators with BLG which is a variant of the model
considered in Ref. \cite{bender}. The generic system is translation invariant, and in order to separate out
the center of mass motion, we consider the following coordinate transformation: 
\bea
x_1 = \frac{x+y}{\sqrt{2}}, \ x_2 = \frac{x - y}{\sqrt{2}}, \ p_1 = \frac{p_x + p_y}{\sqrt{2}},
\ p_2 = \frac{p_x - p_y}{\sqrt{2}}. 
\label{ct1}
\eea
\noindent The Hamiltonian in terms of the new coordinates is expressed as, 
\bea
H & = & p_x^2 - p_y^2 - \gamma ( x p_y + y p_x) - \frac{\gamma^2}{4} ( x^2 - y^2 )  \nonumber \\
  & + & \frac{a}{b} \left\{ e^{b\sqrt{2}y} + e^{-b\sqrt{2}y} -2 \right\}.
\label{ham2}
\eea
\noindent The $\mathcal{P}$ and $\mathcal{T}$ have the standard definition in this coordinate,
namely, $\mathcal{P} : ( x , y ) \rightarrow (x, -y), (p_{x}, p_y) \rightarrow ( p_x, -p_y)$ and
$\mathcal{T} : (x, y) \rightarrow (x, y), (p_x, p_y) \rightarrow (-p_x, -p_y)$.
The Hamiltonian is invariant under the operation of $\mathcal{PT}$. The Hamiltonian is
not positive-definite.

The eigen value equation corresponding to the Hamiltonian (\ref{ham2}) is given as,
\bea
&&\left[  - \left( \partial_x^2 - \partial_y^2 \right) + i \gamma \left( x \partial_y + y \partial_x \right) 
- \frac{\gamma^2}{4} ( x^2 - y^2) \right. \nonumber \\
&& + \left. \frac{a}{b} \left\{ e^{b\sqrt{2}y} + e^{-b\sqrt{2}y} 
- 2 \right\} \right] \psi = \tilde{E} \psi
\label{se1}
\eea
\noindent where we have used the standard coordinate representation  of the Heisenberg equation and
denoted $p_x:=-i \partial_x, \ p_y:=-i \partial_y$. The system is translation invariant and the
associated conserved operator,
\bea
\widehat{\Pi} = \left( - 2i \frac{\partial}{\partial x} + \gamma y \right),
\eea 
\noindent commutes with the Hamiltonian (\ref{ham2}). Thus, the quantum system is integrable as is the case
for the corresponding classical system. 

\begin{figure}[htbp]
\begin{subfigure}{0.47\columnwidth}
          \centering
                \includegraphics[width=0.99\linewidth]{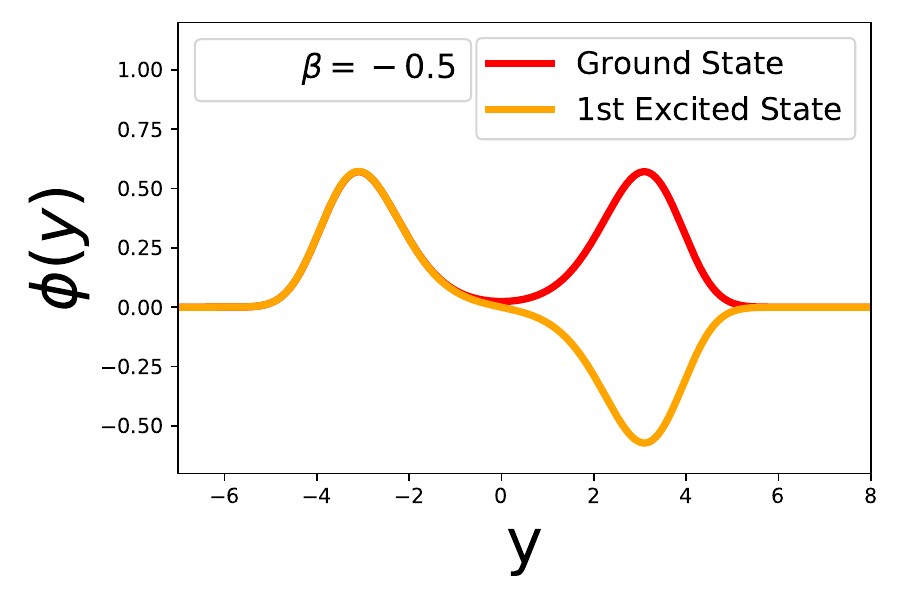}\quad
                \caption{}
                \label{wf1_2d}
        \end{subfigure}
	\begin{subfigure}{0.47\columnwidth}
                \centering
                \includegraphics[width=0.99\linewidth]{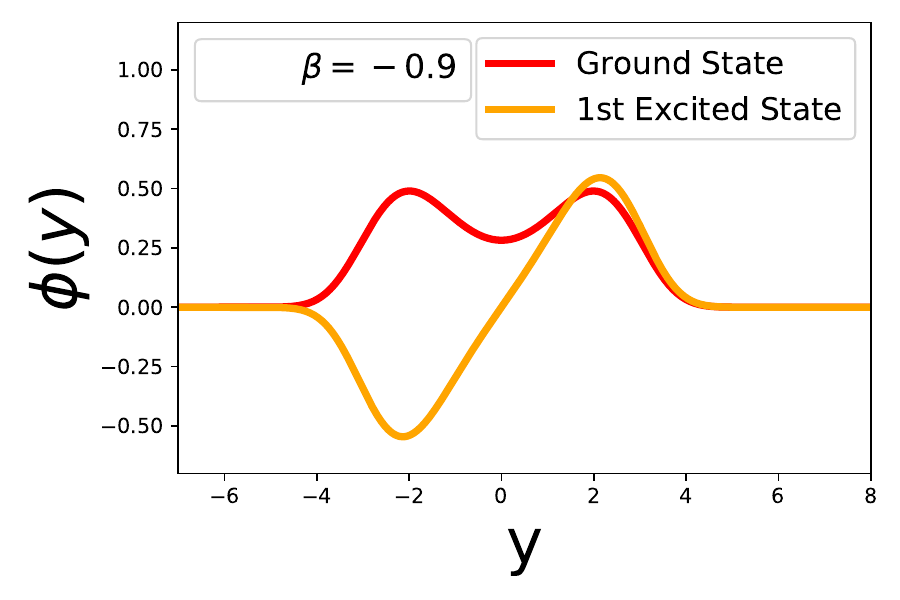}\quad
                \caption{}
                \label{wf2_2d}
        \end{subfigure}
	\medskip
	\begin{subfigure}{0.49\columnwidth}
                \centering
                \includegraphics[width=0.99\linewidth]{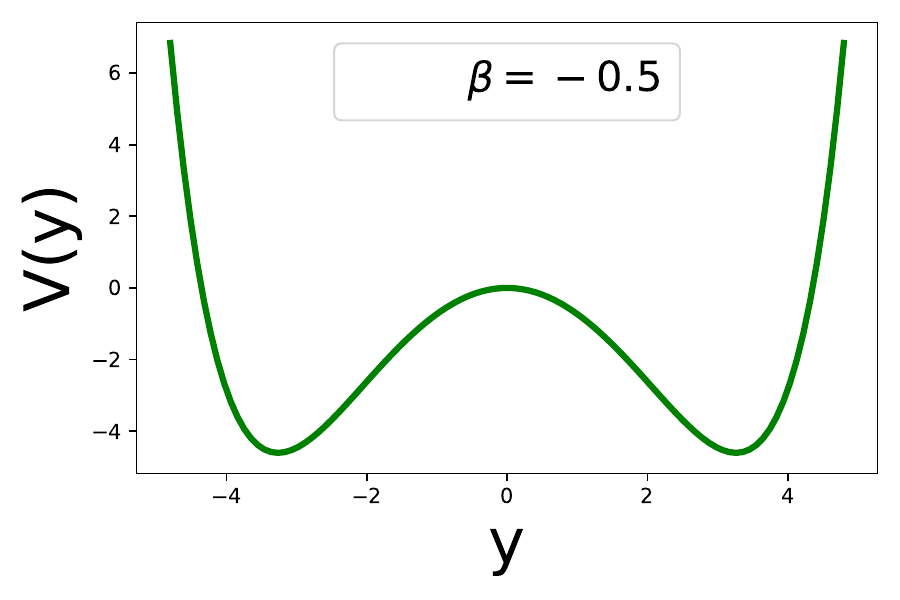}\quad
                \caption{}
                \label{pot_2d_b05}
        \end{subfigure}
        \begin{subfigure}{0.49\columnwidth}
                \centering
                \includegraphics[width=0.99\linewidth]{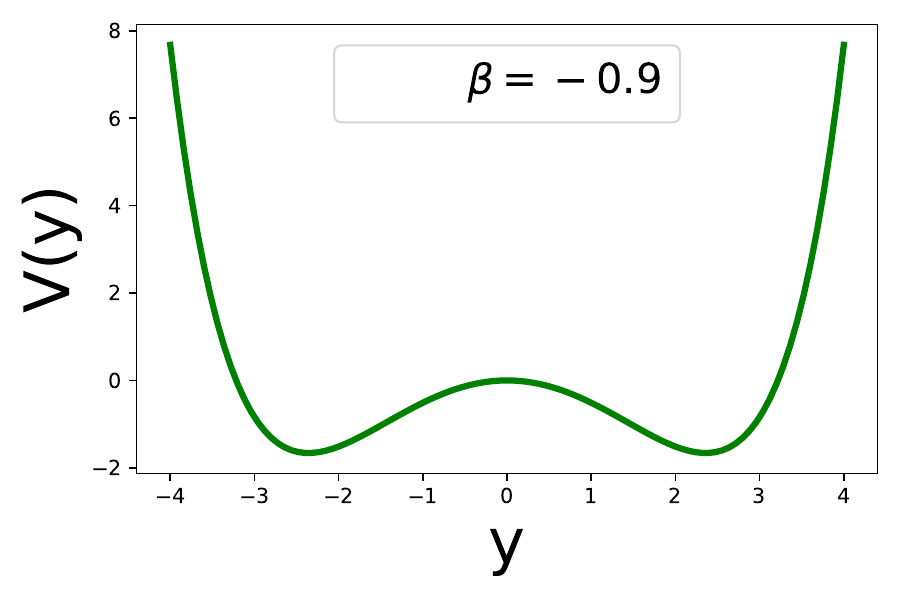}\quad
                \caption{}
		\label{pot_2d_b09}
        \end{subfigure}
	\medskip
	\begin{subfigure}{0.49\columnwidth}
                \centering
                \includegraphics[width=0.99\linewidth]{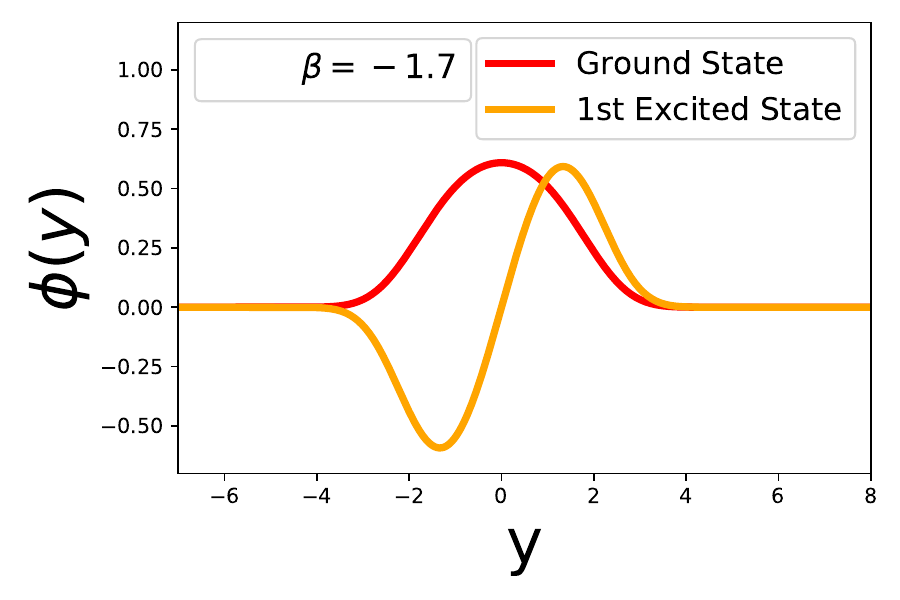}\quad
                \caption{}
		\label{wf3_2d}
        \end{subfigure}
	\begin{subfigure}{0.49\columnwidth}
                \centering
                \includegraphics[width=0.99\linewidth]{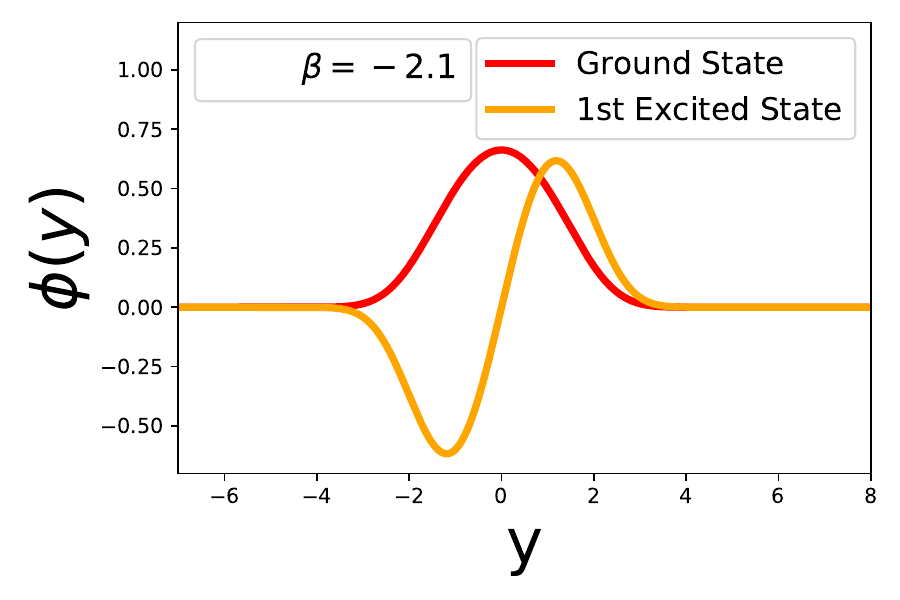}\quad
                \caption{}
                \label{wf4_2d}
        \end{subfigure}
	\medskip
	\begin{subfigure}{0.49\columnwidth}
                \centering
                \includegraphics[width=0.99\linewidth]{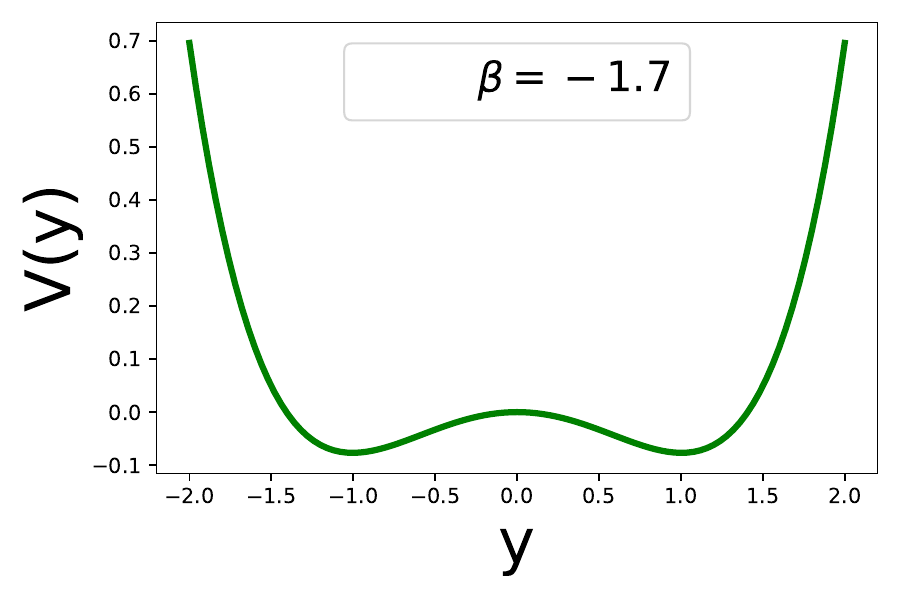}\quad
                \caption{}
                \label{pot_2d_b17}
        \end{subfigure}
        \begin{subfigure}{0.49\columnwidth}
                \centering
                \includegraphics[width=0.99\linewidth]{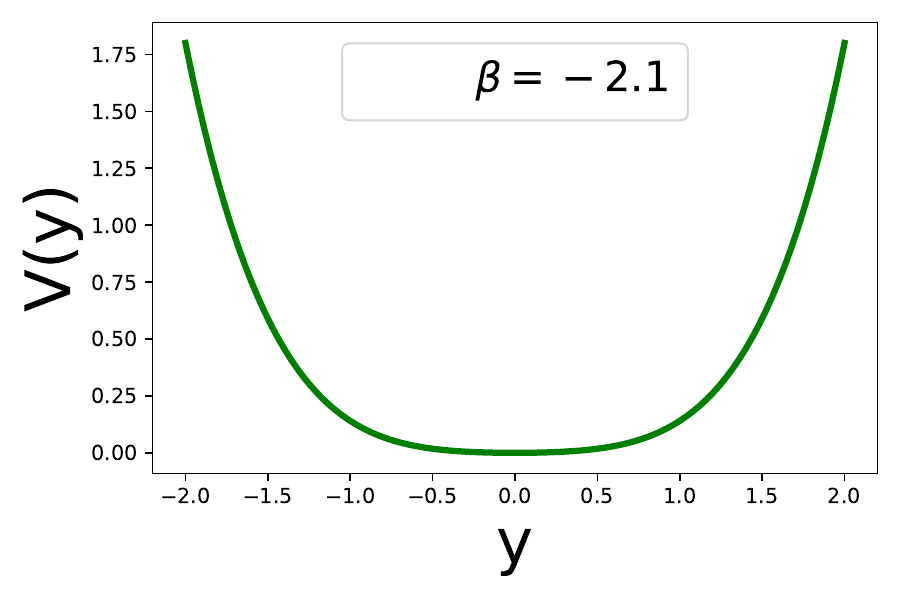}\quad
                \caption{}
                \label{pot_2d_b21}
        \end{subfigure}
	\caption{
	(Color online) Plot of wave functions and potential for different values of $\beta$ and 
	$\lambda=1$,. Green, Red and Orange solid lines represent potential, ground state and 
	1st excited state wave functions, respectively. 
	Fig.(a), Fig.(c) $ \beta = -0.5$; Fig.(b),Fig.(d) $\beta = -0.9$; Fig.(e),Fig.(g)  $\beta = -1.7$; 
	Fig.(f),Fig.(h) $\beta = -2.1$}
	\label{wf_2d}
\end{figure}

The simultaneous wave function of $\widehat{\Pi}$ and $H$ is considered as,
\bea
\psi(x,y) = \exp\left[ \frac{i}{2} x (k - \gamma y) \right] \phi(y),
\label{wf1}
\eea
\noindent which when substituted into Eq. (\ref{se1}) gives the following eigen value 
equation,
\bea
&& \phi^{\prime \prime}(y) + \gamma^2 y^2 \phi(y) + \frac{k}{4} (k - 2 \gamma y) \phi \nonumber \\
&& + \frac{2a}{b} \left\{ \cosh(\sqrt{2}by) - 1\right\} \phi(y) = \tilde{E} \phi(y).
\label{se2}
\eea
\noindent We make a transformation $y \rightarrow y -\frac{k}{4\gamma}$ followed by 
a scale transformation $y \rightarrow \sqrt{2}b y$, and  the resulting eigen value 
equation takes the following form,
\bea
-\phi^{\prime \prime} + \left[ - \lambda y^2  - \beta \left\{ \cosh\left(y
+ y_0 \right) - 1\right\} \right] \phi = E \phi
\label{se4}
\eea
\noindent where $\lambda = \frac{\gamma^2}{4 b^4}$, $ \beta = \frac{a}{b^3}$, 
$E = - \frac{1}{2b^2}(\tilde{E} - \frac{3k^2}{16})$, $y_0 = \frac{bk}{2\sqrt{2}\gamma}$. 
It may be noted that $\lambda$ is always positive and $\beta$ can be both positive as well as negative. 
The eigen value equation in Eq. (\ref{se4}) may be interpreted as that of a particle moving
in the one dimensional potential,
\bea
V(y) = \beta \left(1 - \cosh(y+y_0) \right) - \lambda y^{2}
\eea
The potential $V(y)$ has two localized minima for $-2 \lambda < \beta < 0$ representing a symmetric
double-well potential for $y_0=0$ and an asymmetric one for $y_0 \neq 0$. The potential has only one
minima at $y = 0$, representing a single well, in the remaining region of the parameter-space.
The quantum bound states are allowed only for $\beta < 0$. The potential and wave functions for the
ground and 1st excited states of the system for $\lambda=1, k=0$ and different values of $\beta$ are
given in the Fig. \ref{wf_2d}.
 
\begin{figure}[htbp]
	\centering
	\includegraphics[width = 0.9 \columnwidth]{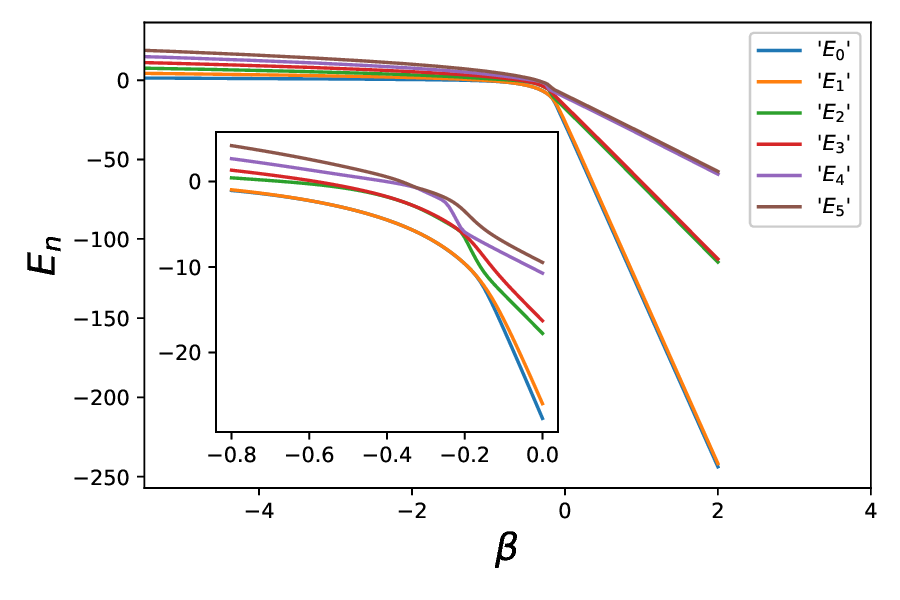}
	\caption{(Color online) Plot of the lowest six energy levels of 2D Toda lattice. 
	Parametric value : $\lambda = 1$}
	\label{level-cross1}
\end{figure}

\begin{figure}[htbp]
        \centering
        \includegraphics[width = 0.85 \columnwidth]{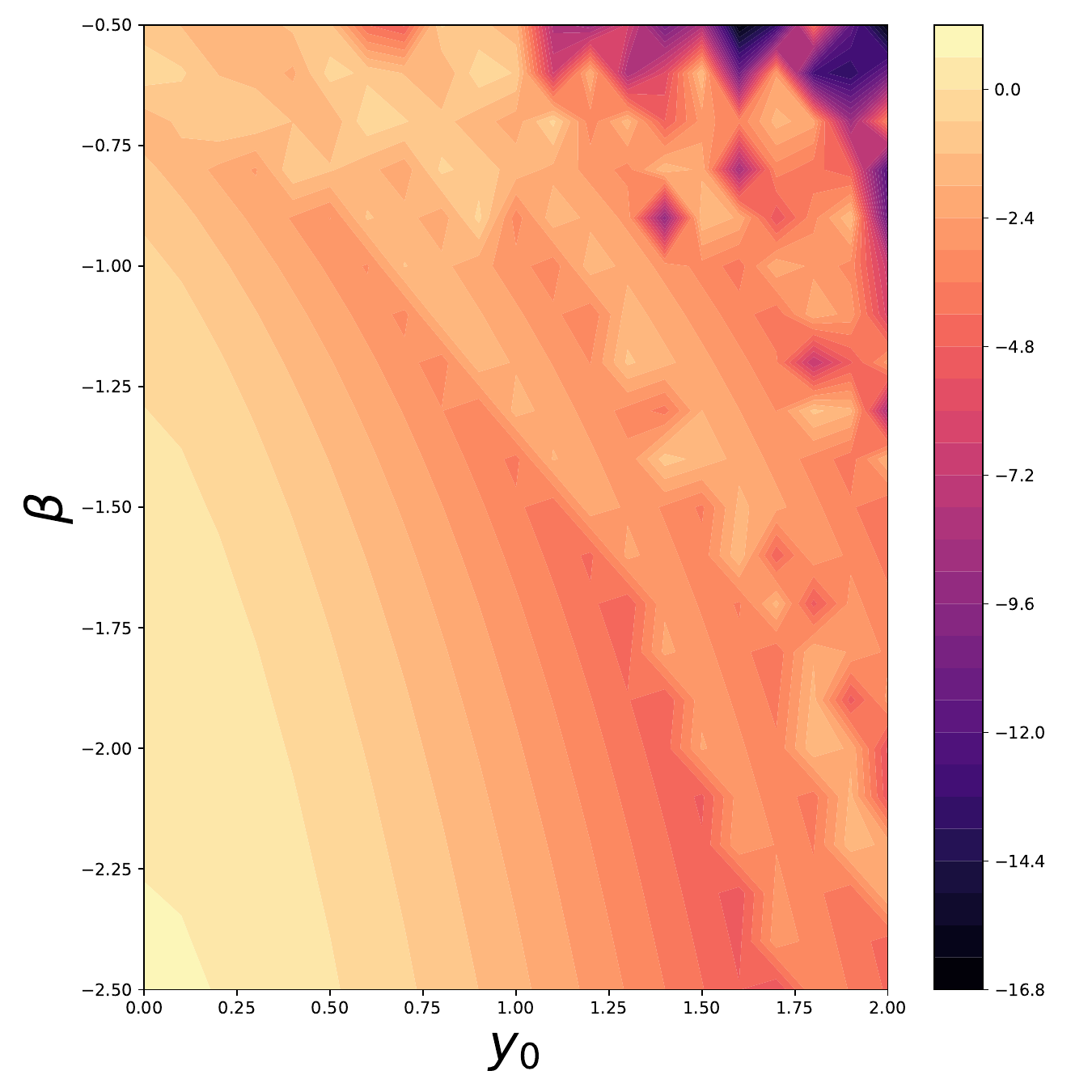}
        \caption{(Color online) Plot of ground state energy of 2D Toda lattice for different 
	values of $\beta$ and $y_0$. Parametric value : $\lambda = 1$}
        \label{2d-spectra}
\end{figure}

\noindent The barrier between the two wells increases with the decreasing value of $|\beta|$, 
and the system decomposes into two independent components separated from each other \textemdash
the wave function is the set of two split up wave functions. The tunnelling probability of the particle
from a well to the other increases with the decrease of the barrier height, and the wave-function
of the left and right regions tend to overlap. The standard behaviour of wave-function
in a double-well represented by a quatric potential is seen in the present context.
The eigen values  corresponding to the quantum bound states of the system is
shown in the Fig. \ref{level-cross1} as a function of $\beta$ for a few low lying states. 
\noindent We observe that the eigen values come close to each other as $\beta$ increases, although the levels 
do not cross each other. The degree of level repulsion is low which is obvious for an integrable 
system. The ground state energy for different values of $\beta$ and $y_0$ is shown in the 
Fig. \ref{2d-spectra}.

\section{Periodic Toda lattice with three particles}

We consider the following Hamiltonian of a Toda lattice with BLG and VMC for three particles,
\bea
H & = & \Pi_1 \Pi_2 + \Pi_1 \Pi_3 + \Pi_2 \Pi_3 + \frac{a}{b} \left[ 1 - e^{b(q_2 - q_1)} \right. \nonumber \\
  & - & \left. e^{b(q_3 - q_2)} - e^{b(q_1 - q_3)} \right]
\eea
\noindent The generalized momenta appearing in $H$ have the following expressions,
\bea
&& \Pi_1 = p_1 + \frac{\gamma}{4} (q_2 + q_3), \ \ \Pi_2 = p_2 - \frac{\gamma}{4} (q_1 - q_3),\nonumber \\
&& \Pi_3 = p_3 - \frac{\gamma}{4} (q_1 + q_2) .
\eea
\noindent The $\gamma$ dependent terms in $\Pi_i$'s produce BLG and VMC in the respective Hamilton's 
equations of motion. In particular, the system is governed by the equations of motion,
\bea
&& \ddot{q}_1 + \gamma \dot{q}_1 + \frac{\gamma}{2} \left ( \dot{q}_2 - \dot{q}_3\right )
-a \left [ e^{b\left(q_2-q_1\right )} -e^{b\left(q_1-q_3\right)}\right]=0\nonumber \\
&& \ddot{q}_2 + \frac{\gamma}{2} \left ( \dot{q}_1- \dot{q}_3 \right )
-a \left [ e^{b\left(q_3-q_2\right )} -e^{b\left(q_2-q_1\right)}\right]=0\nonumber \\
&& \ddot{q}_3 - \gamma \dot{q}_3 + \frac{\gamma}{2} \left ( \dot{q}_1 - \dot{q}_2 \right )
-a \left [ e^{b\left(q_1-q_3\right )} -e^{b\left(q_3-q_2\right)}\right]=0,\nonumber 
\label{toda_eqn}
\eea
\noindent where particle-1 and particle-3 are subjected to BLG. Each particle interacts
with the remaining two particles via the Toda potential and the VMC. 
The above classical system has been studied in detail in Ref. \cite{proy} and shown to
admit mixed phases of integrability and chaos \textemdash two of the integrals of motion
have been obtained analytically. In this article, we study the quantum Hamiltonian
$H$ from the view point of its integrability and chaos.

The system is translation invariant. We introduce the Jacobi coordinates $Q_1,Q_2,Q_3$
in order to separate out the center of mass motion and to work with only two coordinates,
\bea 
Q_{j}  = \frac{1}{\sqrt{j(j+1)}} \sum_{k=1}^{j} ( q_k - j q_{j+1}),
Q_3    =\frac{1}{\sqrt{3}} \sum_{k=1}^{3} q_k,\nonumber 
\eea 
\noindent where $j=1, 2$. The momenta $P_{j}$ in the Jacobi coordinates are related to $p_j$'s through the same relations
\textemdash replace $(Q_k, q_k) \rightarrow (P_k, p_k)$. 
The Hamiltonian can be rewritten in terms of $(Q_k, P_k)$  as,
\bea
H & = & \frac{{\tilde{\Pi}_3}^2}{4} -  \frac{\tilde{\Pi}_1^2}{2} - \frac{\tilde{\Pi}_2^2}{2} 
 - \frac{a}{b} \left[ e^{-b\sqrt{2} Q_1} + e^{\frac{b}{\sqrt{2}} (Q_1 + \sqrt{3} Q_2)} \right. \nonumber \\
  & + & \left. e^{\frac{b}{\sqrt{2}} ( Q_1 - \sqrt{3} Q_2)} - 1 \right]
\label{3dham2}
\eea
\noindent where $\tilde{\Pi}_1$,$\tilde{\Pi}_2$ and $\tilde{\Pi}_3$ are the modified generalized momenta: 
\bea
\tilde{\Pi}_1 & = & - P_1 - \frac{\gamma}{4\sqrt{3}} \left( Q_2 + \sqrt{2} Q_3 \right) \nonumber \\
\tilde{\Pi}_2 & = & - P_2 + \frac{\gamma}{4\sqrt{3}} \left( Q_1 - \sqrt{6} Q_3 \right) \nonumber \\
\tilde{\Pi}_3 & = & 2 P_3 - \frac{\gamma}{\sqrt{6}} \left( Q_1 + \sqrt{3} Q_2 \right)
\eea
\noindent The Hamiltonian is not positive-definite even for $\frac{a}{b} < 0$.

\begin{figure}[htbp]
        \begin{subfigure}{0.49\columnwidth}
                \centering
                \includegraphics[width=0.99\linewidth]{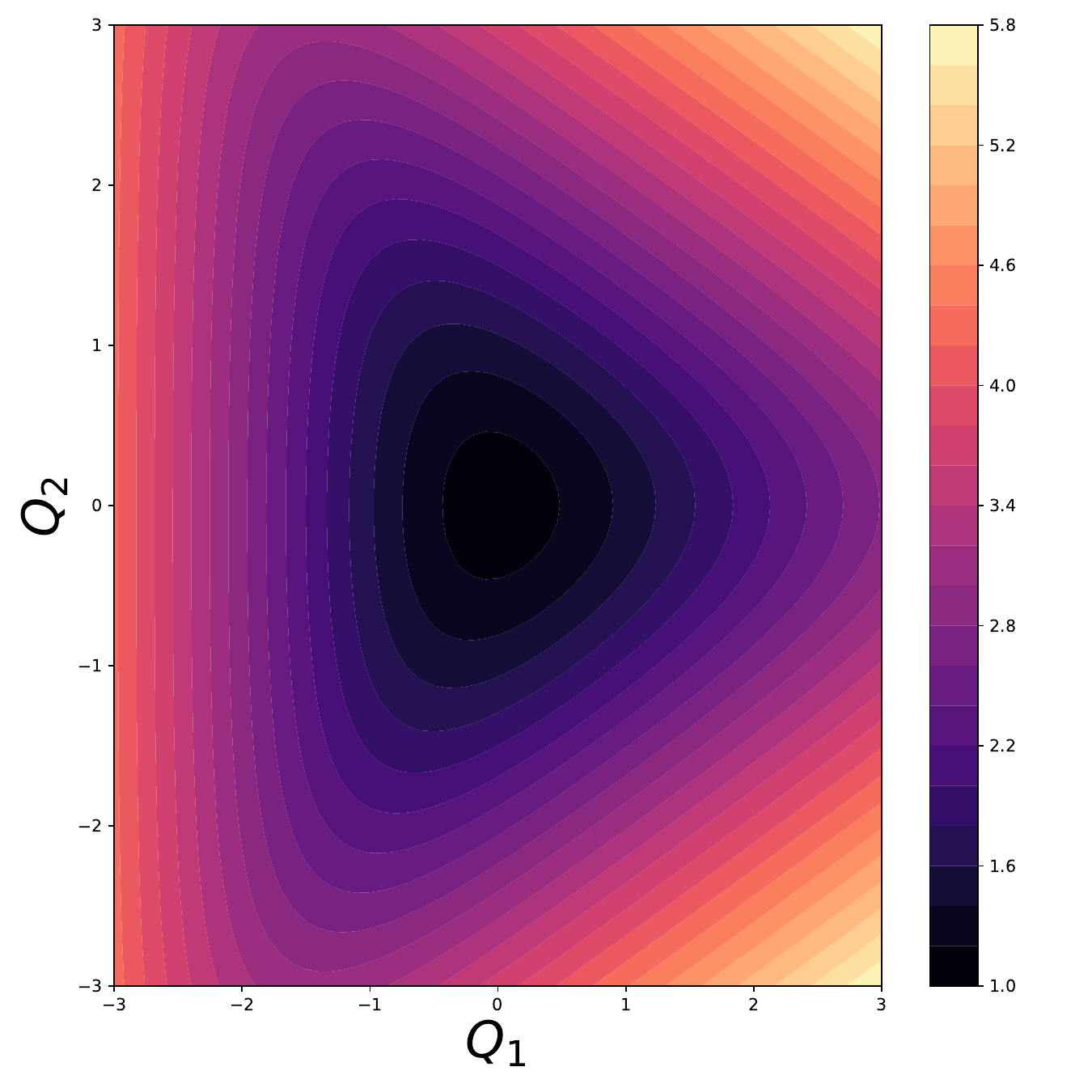}\quad
                \caption{}
                \label{3d_pot1}
        \end{subfigure}
        \begin{subfigure}{0.49\columnwidth}
                \centering
                \includegraphics[width=0.99\linewidth]{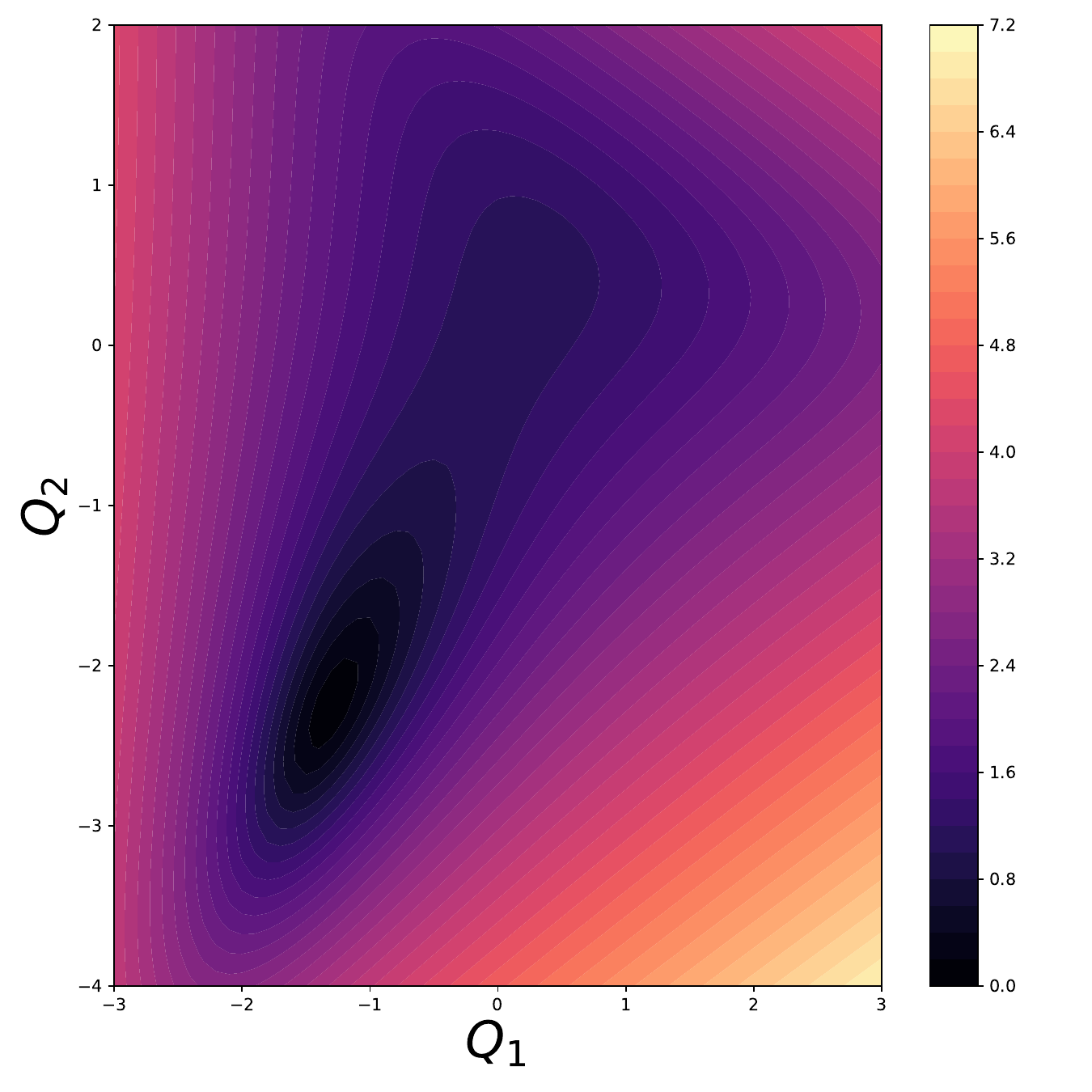}\quad
                \caption{}
                \label{3d_pot2}
        \end{subfigure}
        \medskip
        \begin{subfigure}{0.49\columnwidth}
                \centering
                \includegraphics[width=0.99\linewidth]{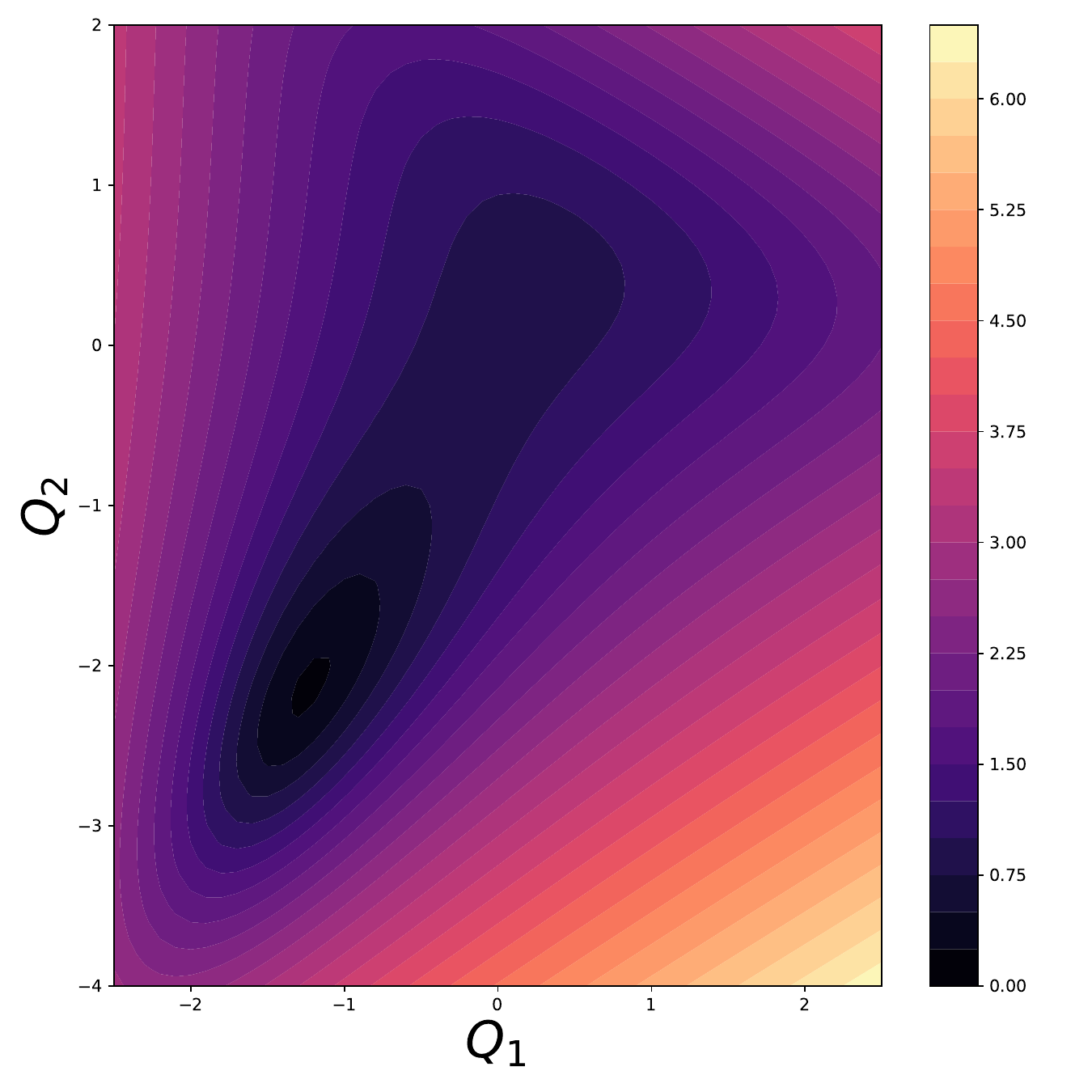}\quad
                \caption{}
                \label{3d_pot3}
        \end{subfigure}
        \begin{subfigure}{0.49\columnwidth}
                \centering
                \includegraphics[width=0.99\linewidth]{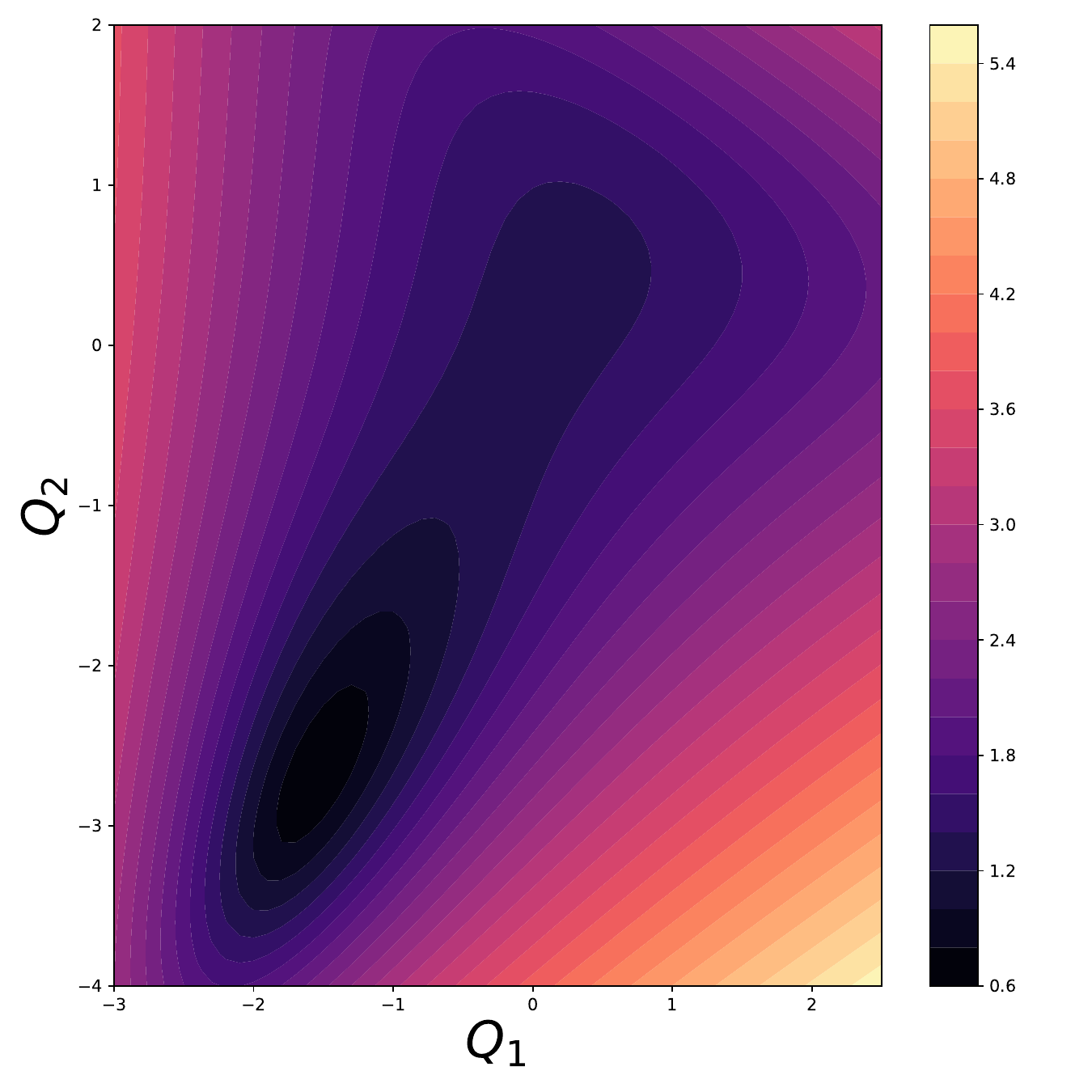}\quad
                \caption{}
                \label{3d_pot4}
        \end{subfigure}
	\caption{(Color online) Contour plot of $log(V_{\textrm{eff}})$ for $k = 0$.
        (a) $a=b=1$,$\gamma = 0.2$; (b) $a=b=1$,$\gamma=1.6$; (c) $a=0.8$,$b=1$,$\gamma=1.4$; 
	(d) $a=1$,$b=0.8$,$\gamma=1.4$}
	\label{3d_potential}
\end{figure}

We quantize the Hamiltonian (\ref{3dham2}) by replacing $(Q_j, P_j)$ with the corresponding operators 
$(Q_j, -i\partial_{Q_j})$ satisfying the commutation relations $[Q_j, P_l] = i \delta_{jl}$, $[Q_j, Q_l] = 0$, 
$[P_j, P_l] = 0$, and denote the resulting Hamiltonian as $\widehat{H}$. We are working with $\hbar=1$.
The translation invariance leads to a conserved quantity and the corresponding operator
\bea
\widehat{\Pi} =  - 2\sqrt{3}i \frac{\partial}{\partial Q_3} + \frac{\gamma}{\sqrt{2}} (Q_1 
+ \sqrt{3} Q_2)
\eea
\noindent  commutes with the Hamiltonian $\widehat{H}$. The complete integrability requires three
integrals of motion. However, we have found analytic expressions for two conserved quantities, namely,
$\widehat{H}$ and $\widehat{\Pi}$. It will be seen later that numerical investigations indicate integrability
of the system in some ranges of the parameter $\gamma$. The commutation relation $[\widehat{H}, \widehat{\Pi}]=0$
allows us to find simultaneous wave function of these two operators,
\bea
\psi(Q_1,Q_2,Q_3) & = & \exp\left[ \frac{i}{2\sqrt{3}} Q_3 \left\{k - \frac{\gamma}{\sqrt{2}} \left(Q_1 
\right. \right. \right. \nonumber \\
                  & + & \left. \left. \left. \sqrt{3} Q_2\right) \right\}  \right] \phi(Q_1,Q_2),
\eea
\noindent where $k$ is the eigen value of the operator $\widehat{\Pi}$. Inserting the expression of 
$\psi(Q_1,Q_2,Q_3)$ into the time-independent schr\"odinger equation 
$\widehat{H} \psi = \epsilon \psi$, we get an eigen value equation in terms of an effective 
Hamiltonian $H_{\textrm{eff}}$ and energy E as $H_{\textrm{eff}} \phi = E \phi$, where $E = - (\epsilon-\frac{a}{b})$. 
The expression of effective Hamiltonian is as follows, 
\bea
H_{\textrm{eff}} & = & \frac{1}{2} \left ( P_1 + \frac{\gamma}{4\sqrt{3}} Q_2 \right )^2 + 
 \frac{1}{2} \left ( P_2 - \frac{\gamma}{4\sqrt{3}} Q_1 \right )^2+ V_{\textrm{eff}}\nonumber \\
V_{\textrm{eff}}& = & - \frac{\gamma^2}{6}
\left ( Q_1 + \sqrt{3} Q_2 +\frac{k}{\sqrt{2} \gamma} \right )^2
+ \frac{a}{b} \left[ e^{-b\sqrt{2} Q_1}
\right. \nonumber \\
& + & \left. e^{\frac{b}{\sqrt{2}} (Q_1 + \sqrt{3} Q_2)} + e^{\frac{b}{\sqrt{2}}
( Q_1 - \sqrt{3} Q_2)} \right],
\label{eff-hami}
\eea
\noindent where $P_1, P_2, Q_1, Q_2$ are to be treated as quantum mechanical operators as mentioned earlier.
The effective Hamiltonian may be interpreted as describing a particle moving in a two-dimensional potential 
$V_{\textrm{eff}}$ and subjected to uniform ``fictitious magnetic field"\cite{ds2,ds3,pkg1} perpendicular to the
``$Q_1-Q_2$"-plane with the absolute magnitude $\frac{|\gamma|}{2\sqrt{3}}$. In the terminology of the
quantization of system with uniform magnetic field, the ``fictitious gauge-potential" in $H_{\textrm{eff}}$ is written
in the symmetric gauge\cite{ds3}. One may choose the Landau gauge and the corresponding
Hamiltonian $H_L$ is obtained through a gauge transformation as,
\bea
H_L & = & exp(- \frac{i \gamma}{4 \sqrt{3}} Q_1 Q_2) \ H_{\textrm{eff}} \ exp( \frac{i \gamma}{4 \sqrt{3}} Q_1 Q_2) \nonumber \\
    & = & \frac{1}{2} (P_1 +\frac{\gamma}{2\sqrt{3}} Q_2)^2 + \frac{P_2^2}{2}  + V_{\textrm{eff}} 
\eea
\noindent A quadratic potential
arises in $V_{\textrm{eff}}$ as the effect of the BLG and VMC, and the standard Toda potential gets
modified. The potential is bounded for $\frac{a}{b} >0$ and contour-plots of the same are shown in Fig. 
(\ref{3d_potential}) for a few choices of the parameters. We study bound states of the quantum Hamiltonian
$H_{\textrm{eff}}$. The analytical treatment of the general eigen value problem appears to be non-trivial,
and will be studied numerically. However, exact solutions may be obtained in the limiting case in which the
Toda potential is approximated as that of coupled oscillators.

\subsection{A Limiting Case}

The Toda potential reduces to that of coupled oscillators in the limit
$b \rightarrow 0, a \rightarrow \infty$ such that $ab \equiv \omega^2$.
The Hamiltonian $H_{\textrm{eff}}$ in this limit reduces to a two-dimensional
anisotropic oscillator in an external uniform magnetic field,
\bea
\tilde{H} & = & \frac{1}{2} \left ( P_1 + \frac{\gamma}{4\sqrt{3}} Q_2 \right )^2 +
\frac{1}{2} \left ( P_2 - \frac{\gamma}{4\sqrt{3}} Q_1 \right )^2\nonumber \\
& + & \frac{9\omega^2 -\gamma^2}{6} Q_1^2 + \frac{3 \omega^2 - \gamma^2}{2} Q_2^2
-\frac{\gamma^2}{\sqrt{3}} Q_1 Q_2,
\eea
\noindent where we have taken $k=0$ and dropped the constant term $\frac{3a}{b}$.
The term $Q_1 Q_2$ can be removed through a rotation by an angle thirty degree in the
anti-clockwise direction. In particular, the transformation
\bea
Q_1 =\frac{\sqrt{3}}{2} X  + \frac{1}{2} Y, \ Q_2 =-\frac{1}{2} X  + \frac{\sqrt{3}}{2} Y
\eea
\noindent transforms $\tilde{H}$ to the following form:
\bea
&& \tilde{H} =  \frac{1}{2} \left ( \Pi_X^2 + \Pi_Y^2 \right ) +
\frac{1}{2} \omega_1^2 X^2 + \frac{1}{2} \omega_2^2 Y^2\nonumber \\
&& \Pi_X= {P_{X} + \frac{\gamma}{4\sqrt{3}} Y, \ \Pi_Y=P_{Y} - \frac{\gamma}{4\sqrt{3}} X}, 
\eea
\noindent where $P_X=-i \partial_x, P_Y=-i \partial_Y$ and $\omega_1^2 =3 \omega^2, \omega_2^2=
\omega_1^2 - \frac{4 \gamma^2}{3}$.  
The Hamiltonian $\tilde{H}$ is $\mathcal{PT}$-symmetric where the parity transformation $\mathcal{P} : 
(X,Y) \rightarrow (X,-Y) , (P_X,P_Y) \rightarrow (P_X,-P_Y)$ and time reversal operation $\mathcal{T} : 
(X,Y) \rightarrow (X,Y) , (P_X,P_Y) \rightarrow (-P_X,-P_Y)$. Phase transition occurs in this systems at 
$9\omega^2 = 4 \gamma^2$.
The eigen values of the systems are real for $\omega^2 > \frac{4}{9} \gamma^2$ and the system is in unbroken 
$\mathcal{PT}$-symmetric phase. The $\mathcal{PT}$-symmetry is broken for $\omega^2 < \frac{4}{9} \gamma^2$ 
and the eigen values become imaginary. 
The Hamiltonian $\tilde{H}$ describes an anisotropic two
dimensional harmonic oscillator in an external uniform magnetic field which has been studied
earlier\cite{gui,rebane}.
We follow the method outlined in Ref. \cite{susy} to obtain the eigenspectra.
Define a four dimensional vector $U=\left ( \omega_1 X, \Pi_X, \omega_2 Y, \Pi_Y \right)$
in the phase-space such that $\tilde{H}=\frac{1}{2} UU^T$, where $U^T$ denotes the transpose
of $U$. It may be noted that $\left [U_i, U_j\right ] = i M_{ij}$ where the $4 \times 4$ matrix
$M$ is given by,
\bea
M = \bp 0 & \omega_1 & 0 & 0\\
-\omega_1 & 0 & 0& \frac{\gamma}{2\sqrt{3}}\\
0&0&0&\omega_2\\ 0& -\frac{\gamma}{2\sqrt{3}}&-\omega_2&0\ep.
\eea
\noindent The eigenvalues of $iM$ are $(\Omega_+,-\Omega_+, \Omega_-,-\Omega_-)$ where
\bea
\Omega_{\pm}= \sqrt{ 3 \left [\omega^2 \pm \frac{5\gamma}{24} \sqrt{\gamma^2+\frac{16}{25} \omega^2}-
\frac{5}{24} \gamma^2 \right ]},
\eea
\noindent and reality of $\Omega_{\pm}$ is ensured for $\omega^2 > \frac{4}{9}
\gamma^2$. The determinant of the matrix $iM$ is zero for $\omega^2 = \frac{4}{9} \gamma^2$
and $\omega = \pm \frac{2}{3} \gamma$ characterizes a critical phase for which $\Omega_{-} = 0$. 
There exists an orthogonal transformation $V=O^T U$ such that the matrix $M$ can be
block-diagonalized as\cite{stopo},
\bea
M_B=O^T M O= \bp 0 & \Omega_+ & 0 & 0\\-\Omega_+ & 0 & 0 & 0\\ 0 & 0 & 0 & \Omega_-\\
0 & 0 & -\Omega_- & 0 \ep .
\eea
\noindent where $O$ is an $O(4)$ rotation matrix and $O^T$ is the transpose of $O$. 
The expression of $O$ is given as,
\begin{widetext}
\bea
O = \bp \frac{\sqrt{2} \omega_1 a_{+}}{\Omega_{+}} ({\omega_2}^2-{\Omega_{+}}^2) & 0 &
\frac{\sqrt{2} \omega_1 a_{-}}{\Omega_{-}} ({\omega_2}^2-{\Omega_{-}}^2) & 0 \\
	0 & \sqrt{2} a_{+} ({\omega_2}^2-{\Omega_{+}}^2) & 0 & \sqrt{2} a_{-} ({\omega_2}^2-{\Omega_{-}}^2) \\
0 & -\frac{\omega_2 \gamma a_{+}}{\sqrt{6}} & 0 & -\frac{\omega_2 \gamma a_{-}}{\sqrt{6}} \\
\frac{\gamma a_{+} \Omega_{+}}{\sqrt{6}} & 0 & \frac{\gamma a_{-} \Omega_{-}}{\sqrt{6}} & 0 \ep,
\eea
\end{widetext}
\noindent where 
\bea
a_{\pm} = \frac{\Omega_{\pm}}{\sqrt{(\omega_2^2 - {\Omega_{\pm}}^2)^2({\omega_1}^2 + {\Omega_{\pm}}^2) 
+ \frac{\gamma^2 {\Omega_{\pm}}^2}{12} ({\omega_2}^2 + {\Omega_{\pm}}^2)}} \nonumber .
\eea
\noindent The matrix $O$ is unique modulo $O(4)$ rotations. 
We denote $M_D=\textrm{diag}(\Omega_+, - \Omega_+,\Omega_-, -\Omega_-)$ and let $S$ diagonalizes 
$iM$, i. e. $M_D= S^{\dagger} (iM) S$. The similarity transformation can not change eigenvalues 
and let $T$ diagonalizes $i M_B$, i.e. $M_D=T^{\dagger} (i M_B) T$. The matrix $O$ is constructed 
as $O=ST^{\dagger}$. 
The transformed variables in the phase space $(x, p_x, y, p_y) \equiv ( v_1/{\sqrt{\Omega_+}},
v_2/{\sqrt{\Omega_+}}, v_3/{\sqrt{\Omega_-}}, v_4/{\sqrt{\Omega_-}})$ where $\Omega_{-} \neq 0$, satisfy the
standard commutation relations,
\bea
[x,p_x]=i, \ [y,p_y]=i, \ [x,y]=0=[p_x,p_y].
\eea
\noindent The Hamiltonian is expressed in terms of new variables as two decoupled
anisotropic harmonic oscillators,
\bea
\tilde{H}= \frac{\Omega_+}{2} \left (p_x^2 +x^2\right ) +\frac{\Omega_-}{2} \left (p_y^2 + y^2 \right ).
\eea
\noindent The Hamiltonian can be expressed in terms of two sets of annihilation and creation operators as,
\bea
&& \tilde{H}=\Omega_+ \left ( a^{\dagger} a +\frac{1}{2} \right )
+ \Omega_- \left ( b^{\dagger} b +\frac{1}{2} \right ) ,\nonumber \\
&& a=\frac{1}{\sqrt{2}} \left  ( p_x - i x \right ),\ \
b=\frac{1}{\sqrt{2}} \left  ( p_y - i y \right )
\eea
\noindent The ground state is determined from the conditions $a \Psi_0=0, b \Psi_0=0$,
\bea
\Psi_0(x,y)= \frac{2}{\sqrt{\pi}} \ e^{-\frac{1}{2} \left (x^2 + y^2\right )}, \ \
E_{0,0}=\frac{\Omega_++\Omega_-}{2}.
\eea
\noindent The energy eigenvalues and the eigenfunctions are,
\bea
&& E_{n,m} = \left ( n +\frac{1}{2} \right ) \Omega_+ + \left ( m+
\frac{1}{2} \right ) \Omega_-, \ n,m \in \mathbb{Z}^{\geq 0}\nonumber \\
&& \Psi_{n,m}(x,y) = \frac{(a^{\dagger})^n}{\sqrt{n!}} \frac{(b^{\dagger})^m}{\sqrt{m!}} \Psi_0(x,y).
\eea
\noindent The wave-function $\Psi_{n,m}(x,y)$ may be expressed in terms of the original variables through
a series of inverse transformations which we do not pursue here. It may be noted that there is
a reduction in the phase-space for $\omega^2=\frac{4}{9}\gamma^2$ for which $\Omega_-=0$ and the
eigenvalues become infinitely degenerate. The energy eigen values are complex for
$\omega^2 < \frac{4}{9}\gamma^2$ and entirely real for $\omega^2 > \frac{4}{9}\gamma^2$.

\subsection{General Hamiltonian}

The eigen value problem for the Hamiltonian $H_{eff}$ in its full generality seems to be unamenable
for an analytical treatment. We study the  quantum bound states of $H_{eff}$ numerically, and obtain the
energy spectra and wave functions by using finite difference method. The scattering
states of $H_{eff}$, if any, have been excluded from our purview of study.
We have considered $\gamma >0, k \geq 0$ and $a=b=1$ in the numerical calculation unless otherwise
stated explicitly. The eigen values and eigenfunctions thus obtained are analyzed to investigate quantum
integrability and chaos in the system. In particular, we study avoid level crossing, probability distribution
of wave functions in the semi-classical region, level statistics and gap-ratio distribution  in order
to study quantum integrability and chaos. It is worth recalling that the system in Ref. \cite{proy}
is integrable for $|\gamma| \lessapprox 1.5$ and chaotic above this critical value. We present numerical
results in the following for $\gamma$ varying from zero to values above this critical in order to search
for quantum integrable and chaotic region. 

\subsubsection{Avoided level crossing and probability distribution}

\begin{figure}[htbp]
	\centering
	\includegraphics[width=0.98\columnwidth]{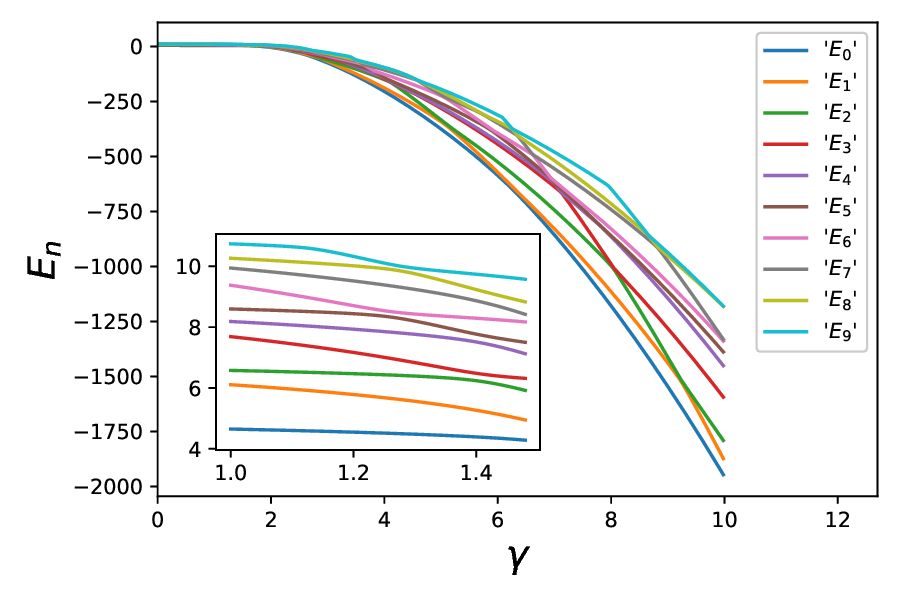}
	\caption{(Color online) Plot of lowest ten energy eigen value of 3-particle system 
	as a function of BLG parameter $\gamma$ for $k=0$ and $a=b=1$}
	\label{level-cross2}
\end{figure}

\begin{figure}[htbp]
        \centering
        \includegraphics[width=0.95\columnwidth]{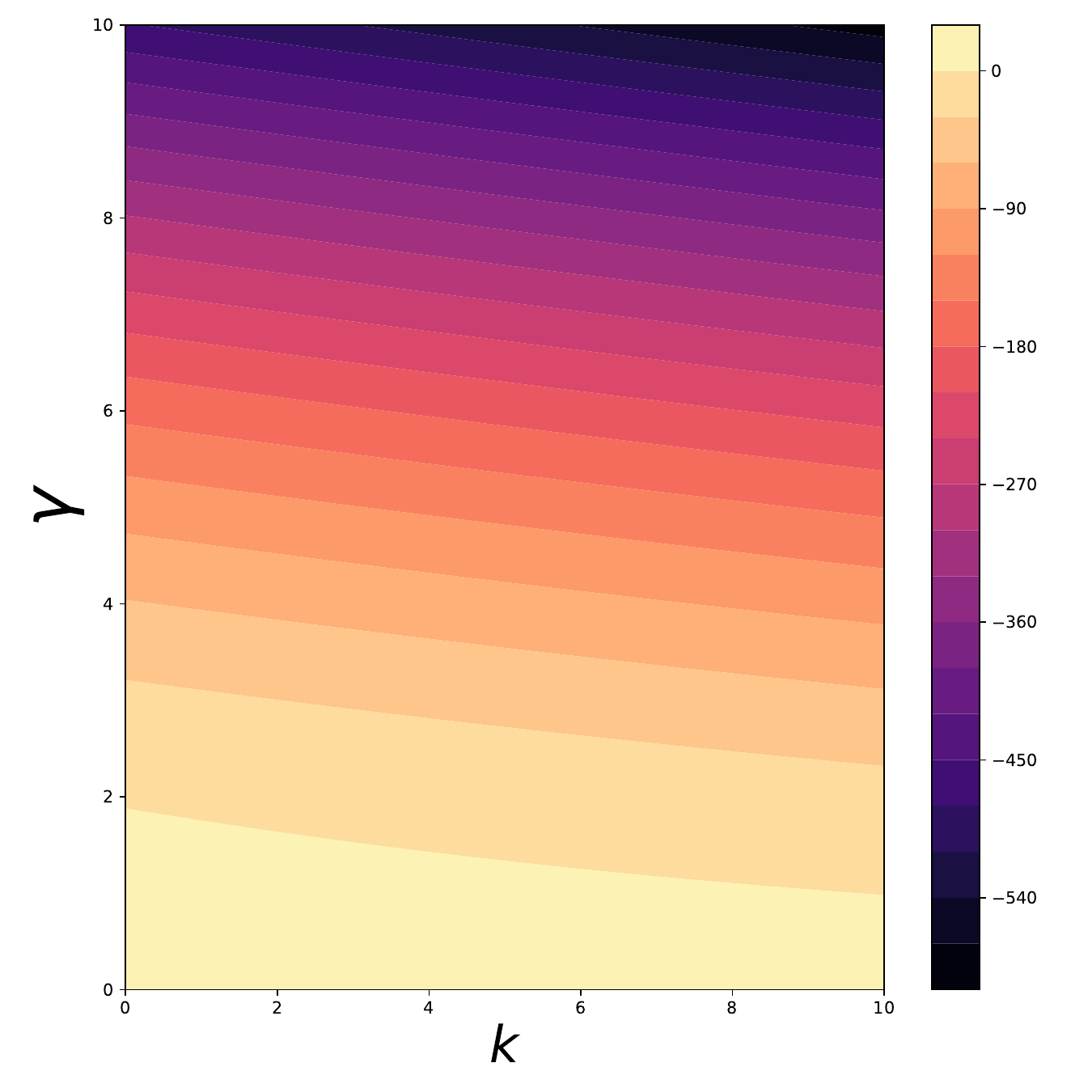}
        \caption{(Color online) Plot of ground state energy for different values of 
	$\gamma$ and $k$. Here $a=b=1$}
	\label{3d-spectra}
\end{figure}

The lowest ten energy eigen values of $H_{\textrm{eff}}$ for different values of $\gamma$, and
$k=0, a=b=1$ are shown in Fig. \ref{level-cross2}. The energies corresponding to each  state decrease with
increasing $\gamma$. The energy levels are highly correlated and repel each other.
The avoided level crossing is a signature of quantum chaos\cite{haake}, since level-crossing
gives rise to the degeneracy in the spectrum, implying symmetry and associated conserved quantities
in the system. The avoided level crossing is seen for larger values
of $\gamma$ in Fig. \ref{level-cross2}. The level repulsion is not obvious from the
figure for smaller values of $\gamma$, and is shown separately
in the inset of Fig. \ref{level-cross2} for $1 \leq \gamma \leq 1.5$.
The dependence of the ground state energy of  $H_{\textrm{eff}}+ \frac{k^2}{12}$ on $\gamma$ and
$k$  is shown in Fig. \ref{3d-spectra}. The ground state energy is shifted for $k \neq 0$ by
an amount of the order of $-\gamma^2 k$ and the same behaviour is seen.
 

\begin{figure}[htbp]
	\begin{subfigure}{0.49\columnwidth}
		\centering
		\includegraphics[width=0.99\linewidth]{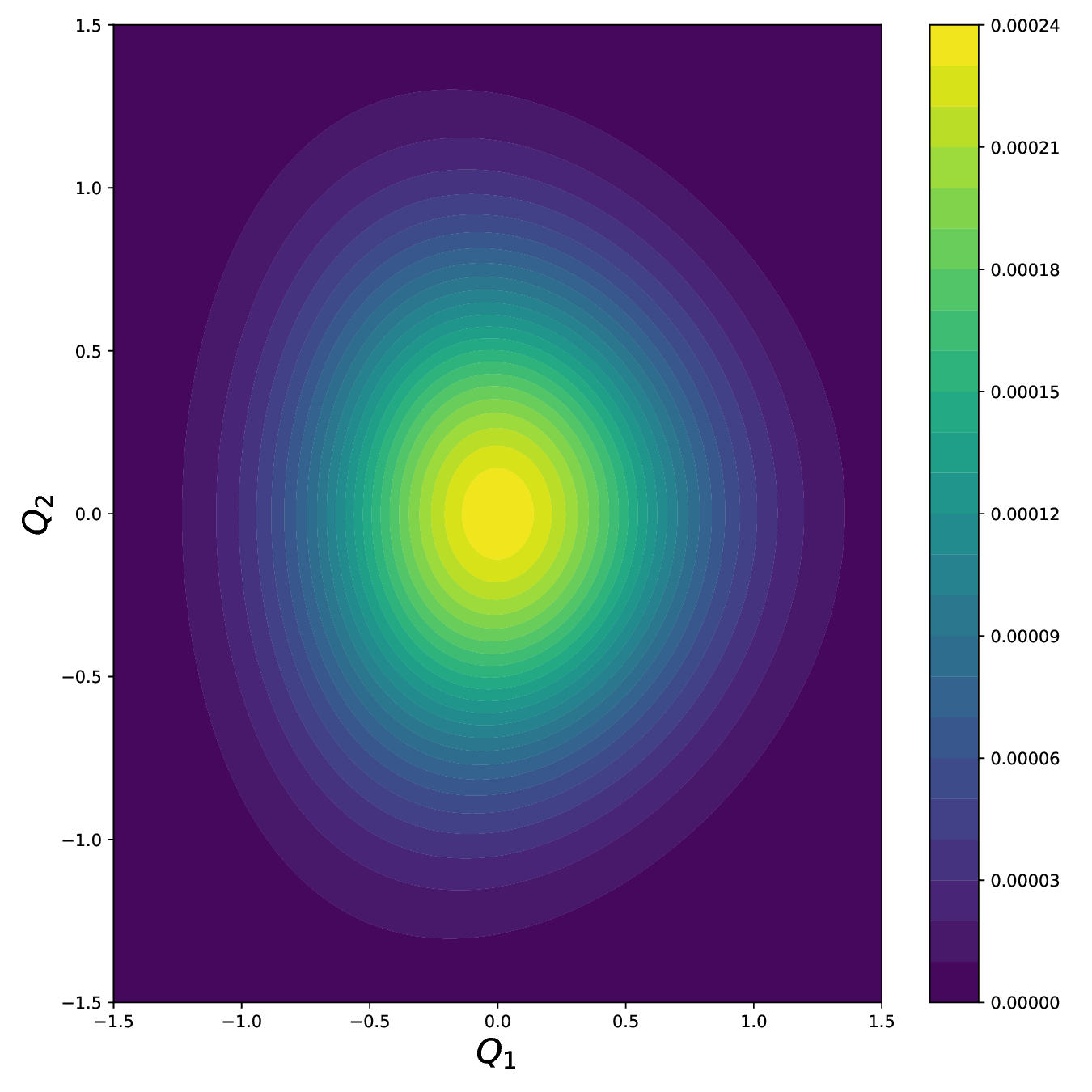}\quad
		\caption{}
		\label{3dwf1}
	\end{subfigure}
	\begin{subfigure}{0.49\columnwidth}
		\centering
		\includegraphics[width=0.99\linewidth]{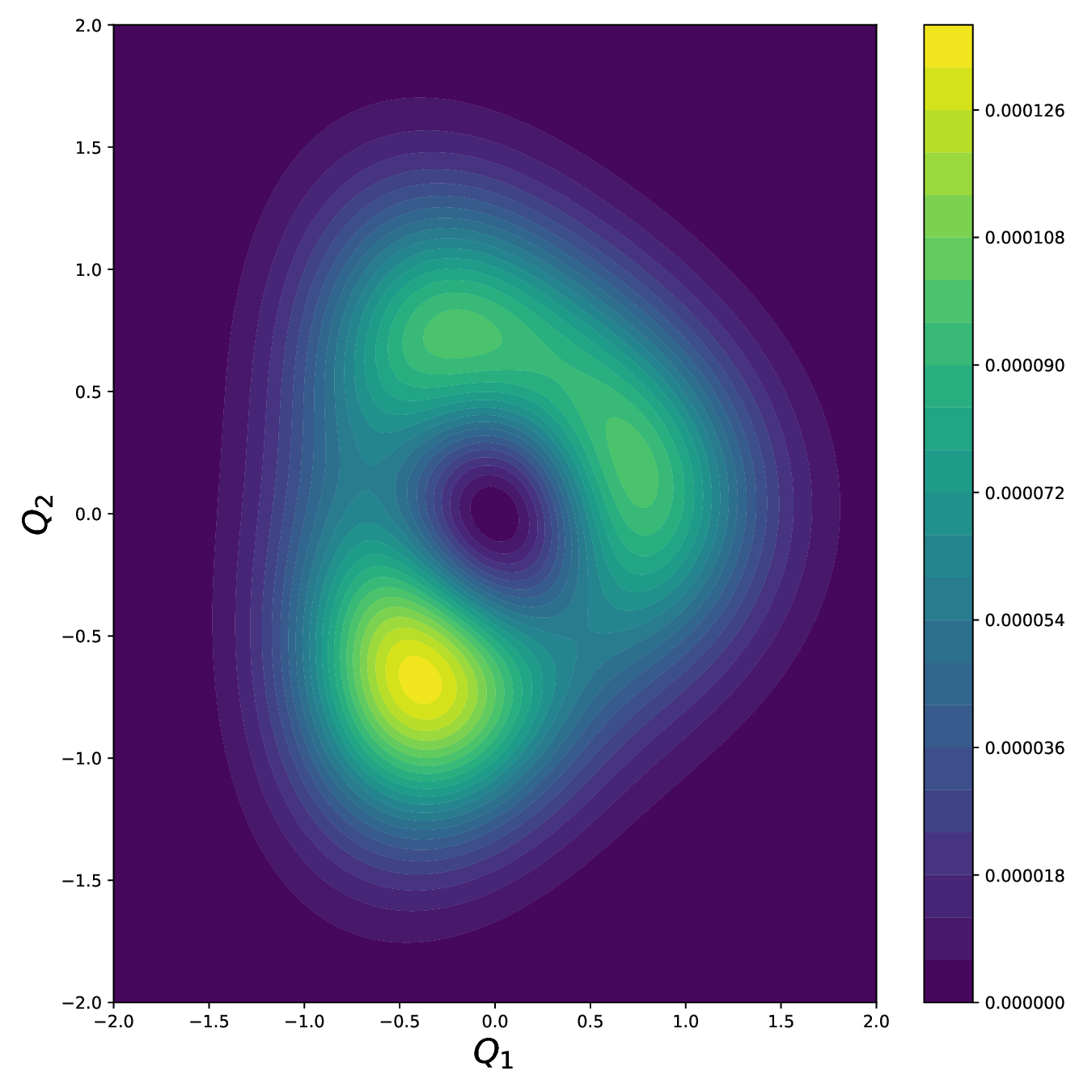}\quad
		\caption{}
		\label{3dwf2}
	\end{subfigure}
	\medskip
	\begin{subfigure}{0.49\columnwidth}
		\centering
		\includegraphics[width=0.99\linewidth]{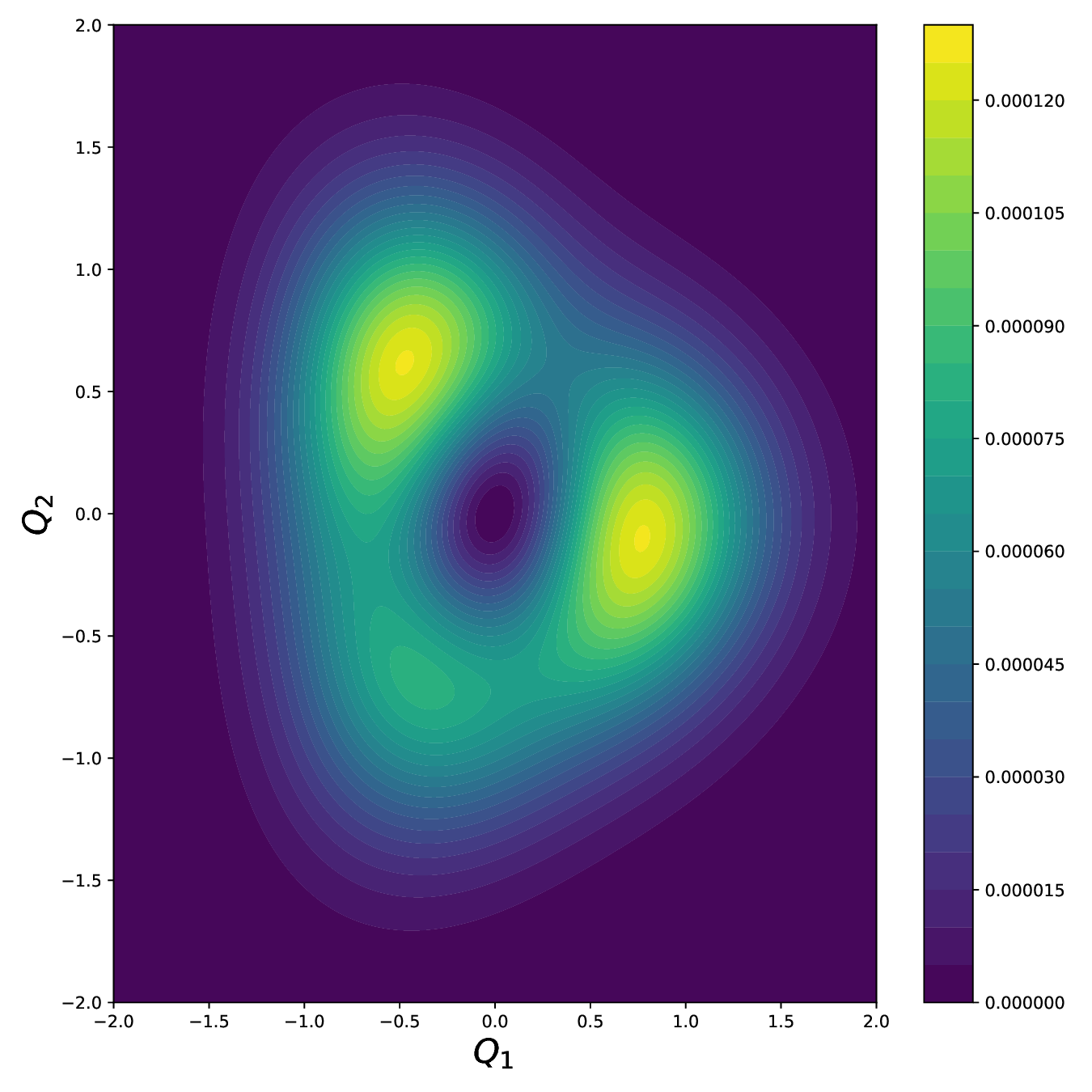}\quad
		\caption{}
		\label{3dwf3}
	\end{subfigure}
        \begin{subfigure}{0.49\columnwidth}
                \centering
                \includegraphics[width=0.99\linewidth]{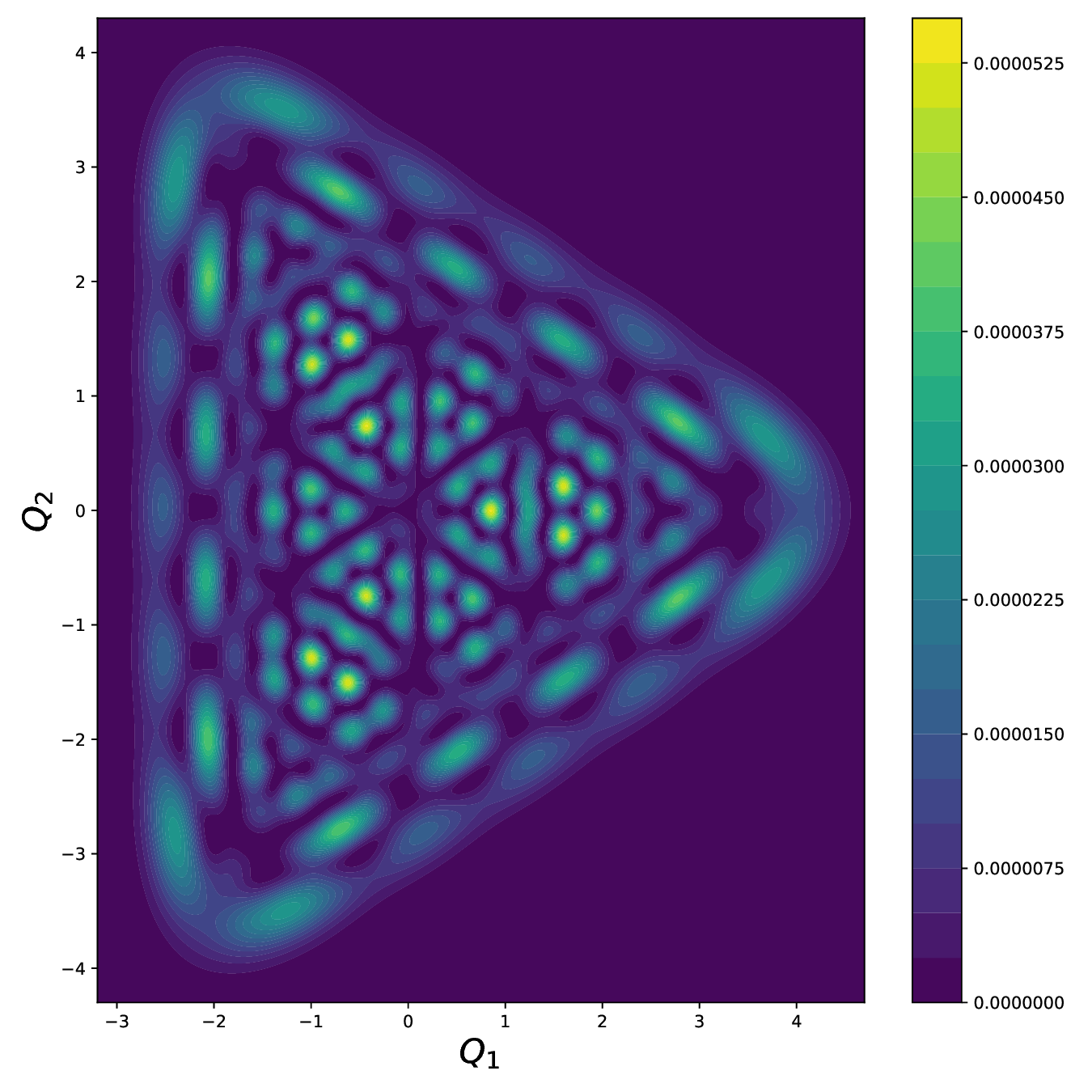}\quad
                \caption{}
                \label{3dwf6}
        \end{subfigure}
	\medskip
        \begin{subfigure}{0.49\columnwidth}
                \centering
                \includegraphics[width=0.99\linewidth]{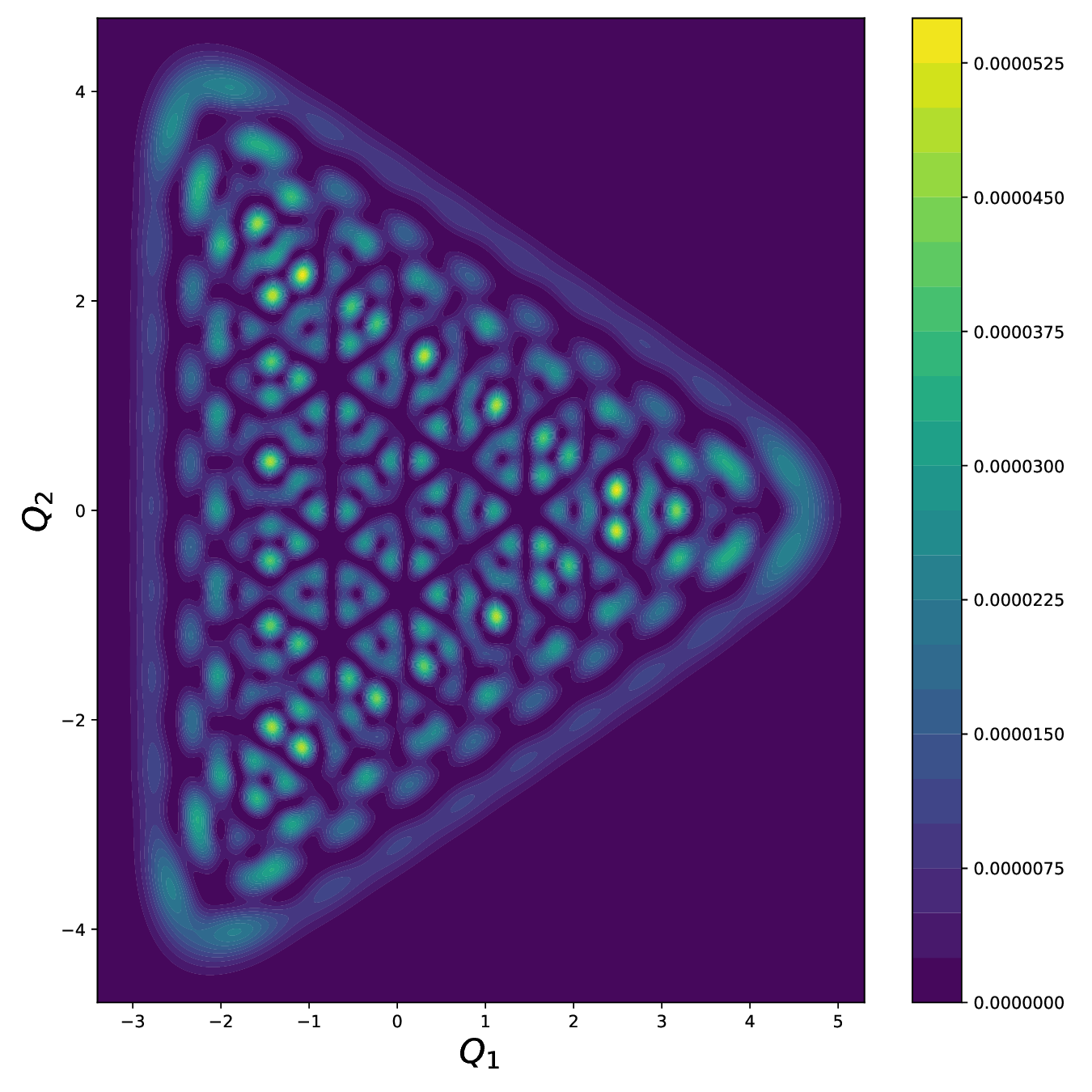}\quad
                \caption{}
                \label{3dwf7}
        \end{subfigure}
        \begin{subfigure}{0.49\columnwidth}
                \centering
                \includegraphics[width=0.99\linewidth]{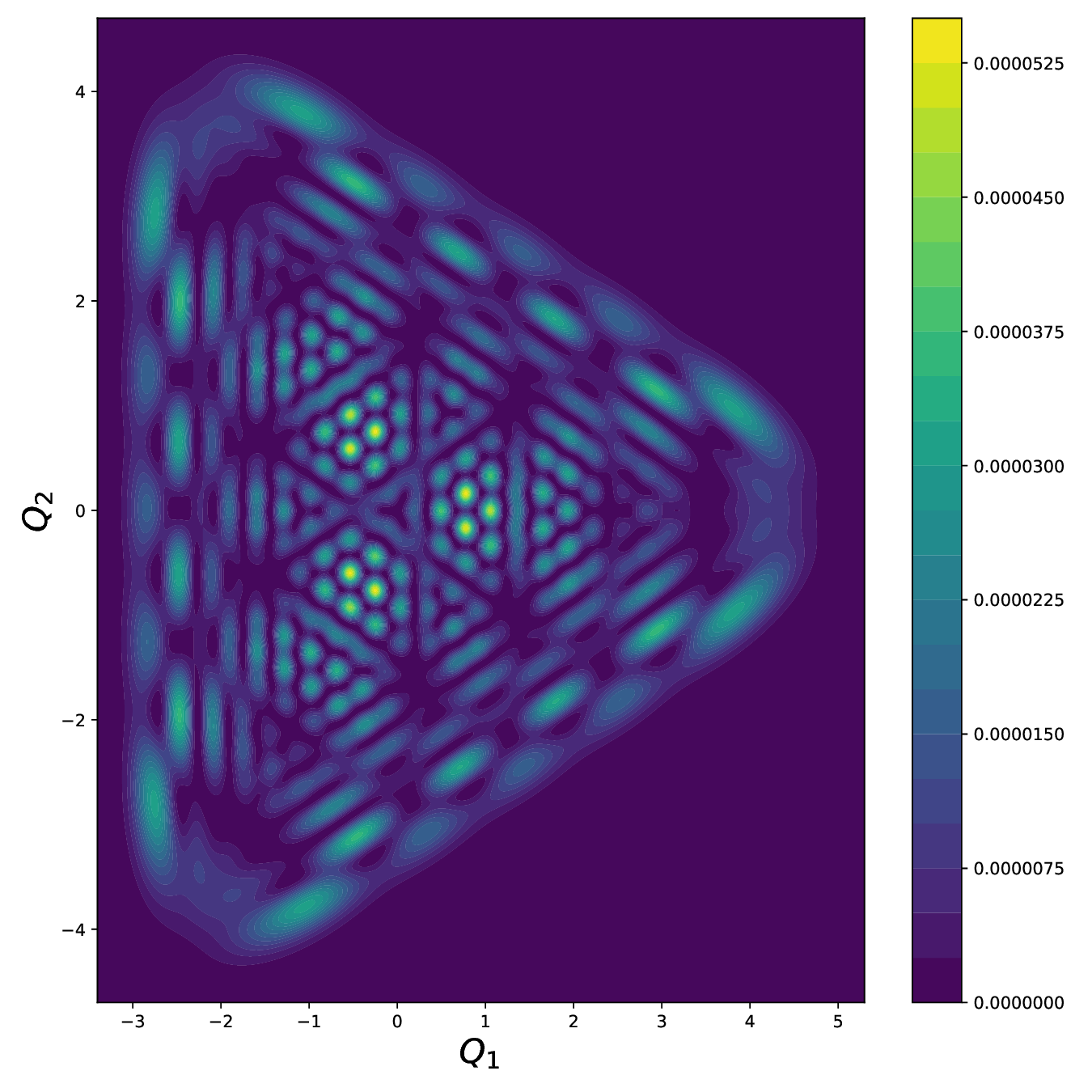}\quad
                \caption{}
                \label{3dwf8}
        \end{subfigure}
	\caption{(Color online) Plot of $|\Phi|^{2}$ for $k = 0$,$a=b=1$ and $\gamma = 0.2$. 
	(a) Ground state; (b) 1st excited state; (c) 2nd excited state; (d) 
	200th excited state; (e) 300th excited state; (f) 400th excited state.}
	\label{wf-integrable}
\end{figure}

\begin{figure}[htbp]
        \begin{subfigure}{0.49\columnwidth}
                \centering
		\includegraphics[width=0.99\linewidth]{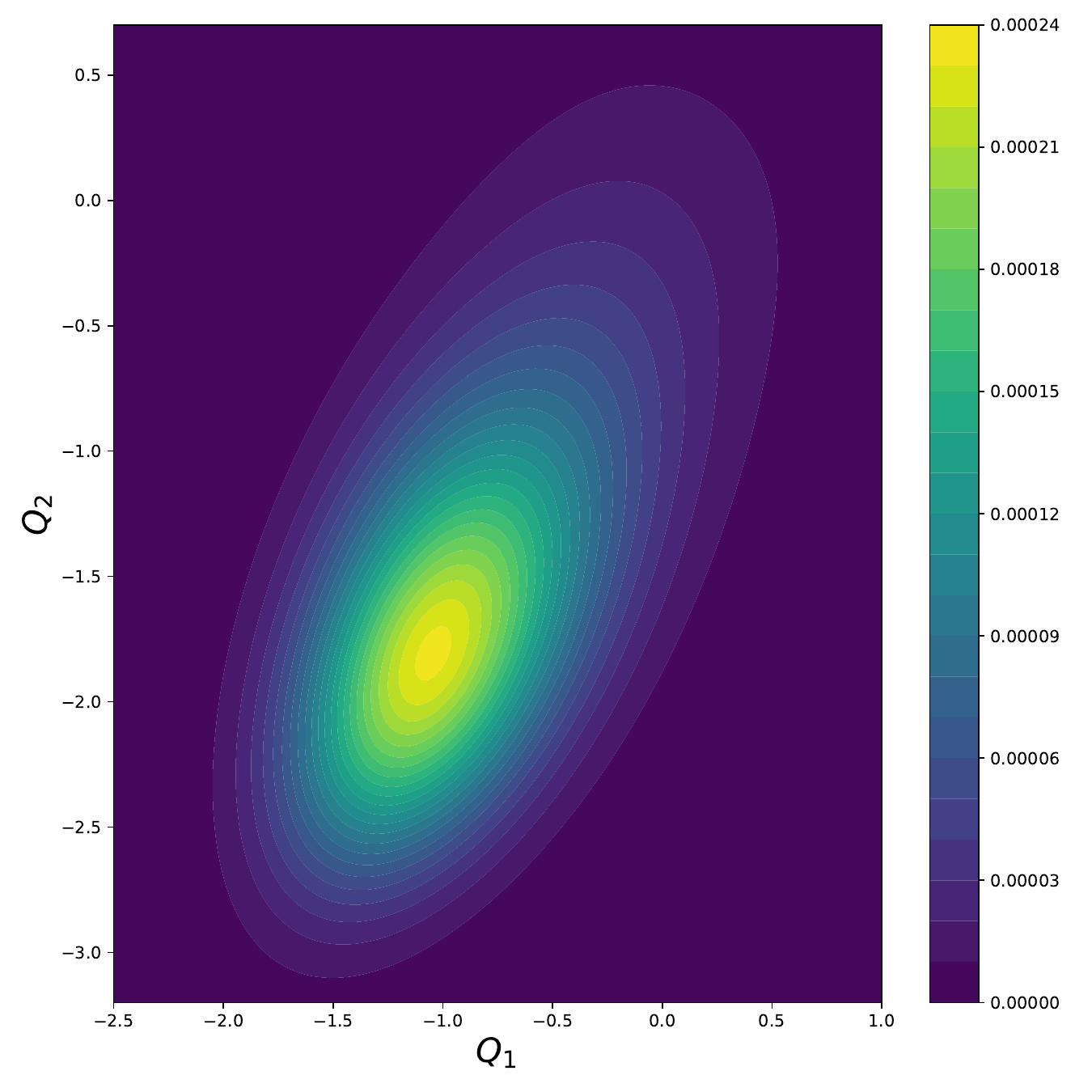}\quad
                \caption{}
                \label{ch_wf1}
        \end{subfigure}
        \begin{subfigure}{0.49\columnwidth}
                \centering
                \includegraphics[width=0.99\linewidth]{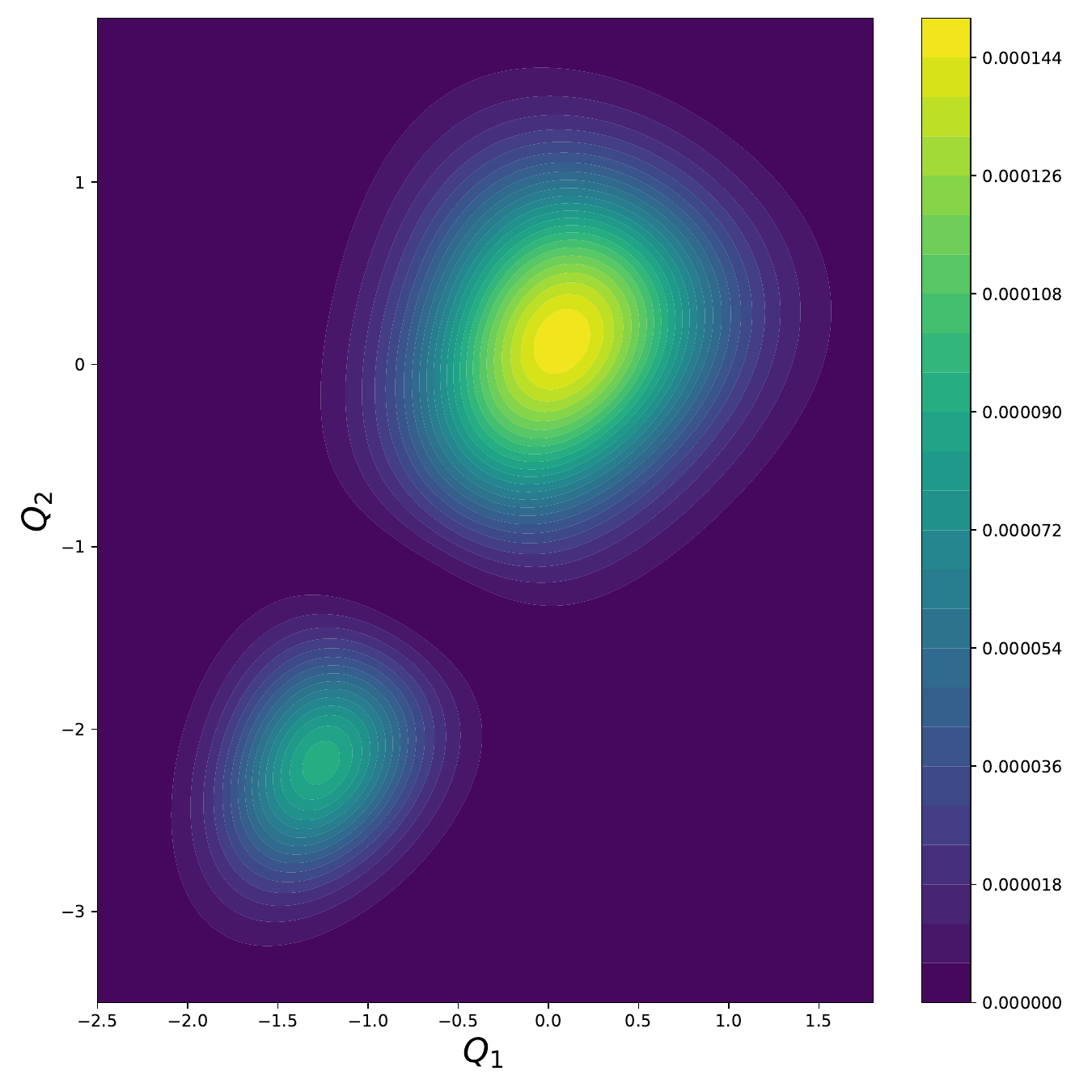}\quad
                \caption{}
                \label{ch_wf2}
        \end{subfigure}
	\medskip
        \begin{subfigure}{0.49\columnwidth}
                \centering
                \includegraphics[width=0.99\linewidth]{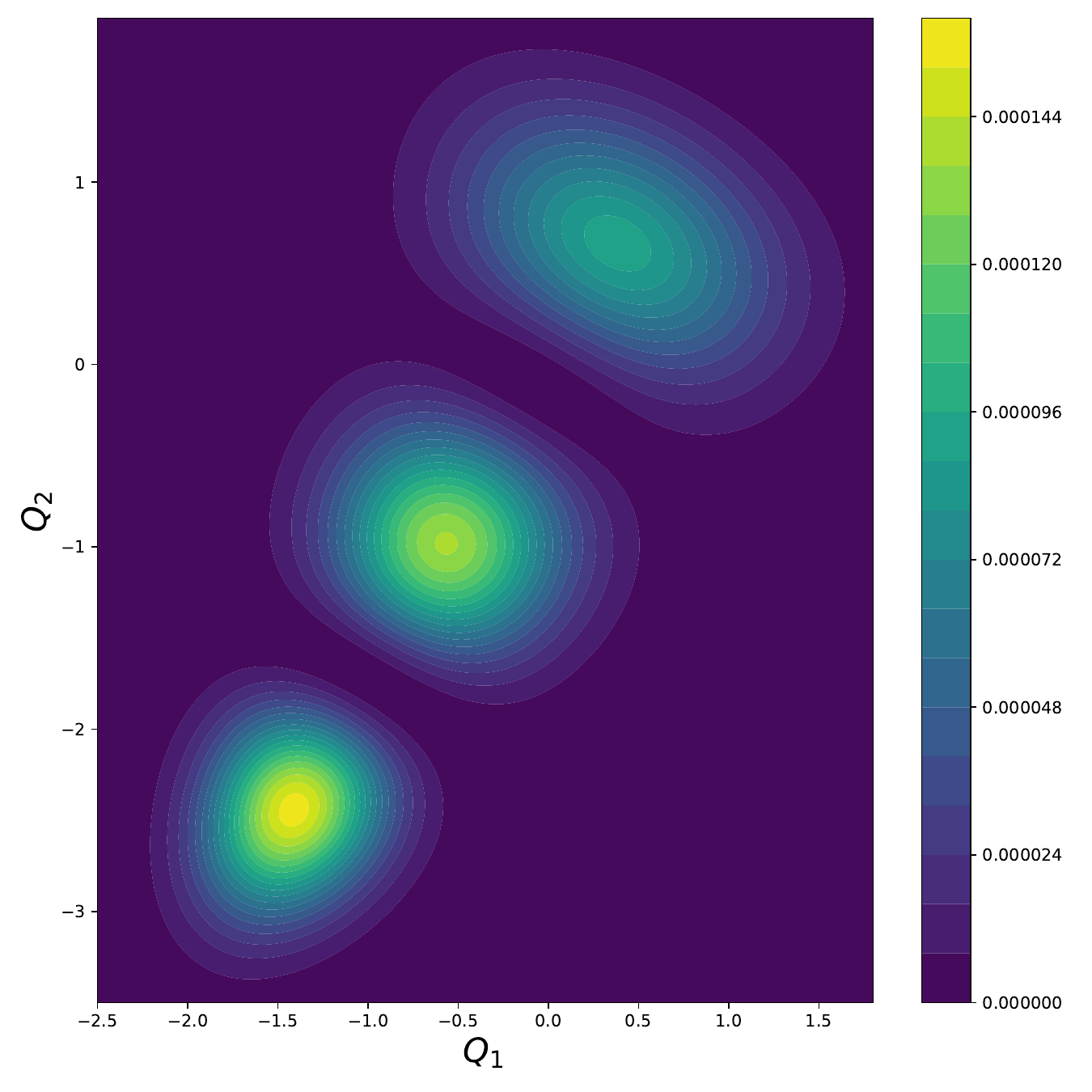}\quad
                \caption{}
                \label{ch_wf3}
        \end{subfigure}
        \begin{subfigure}{0.49\columnwidth}
                \centering
                \includegraphics[width=0.99\linewidth]{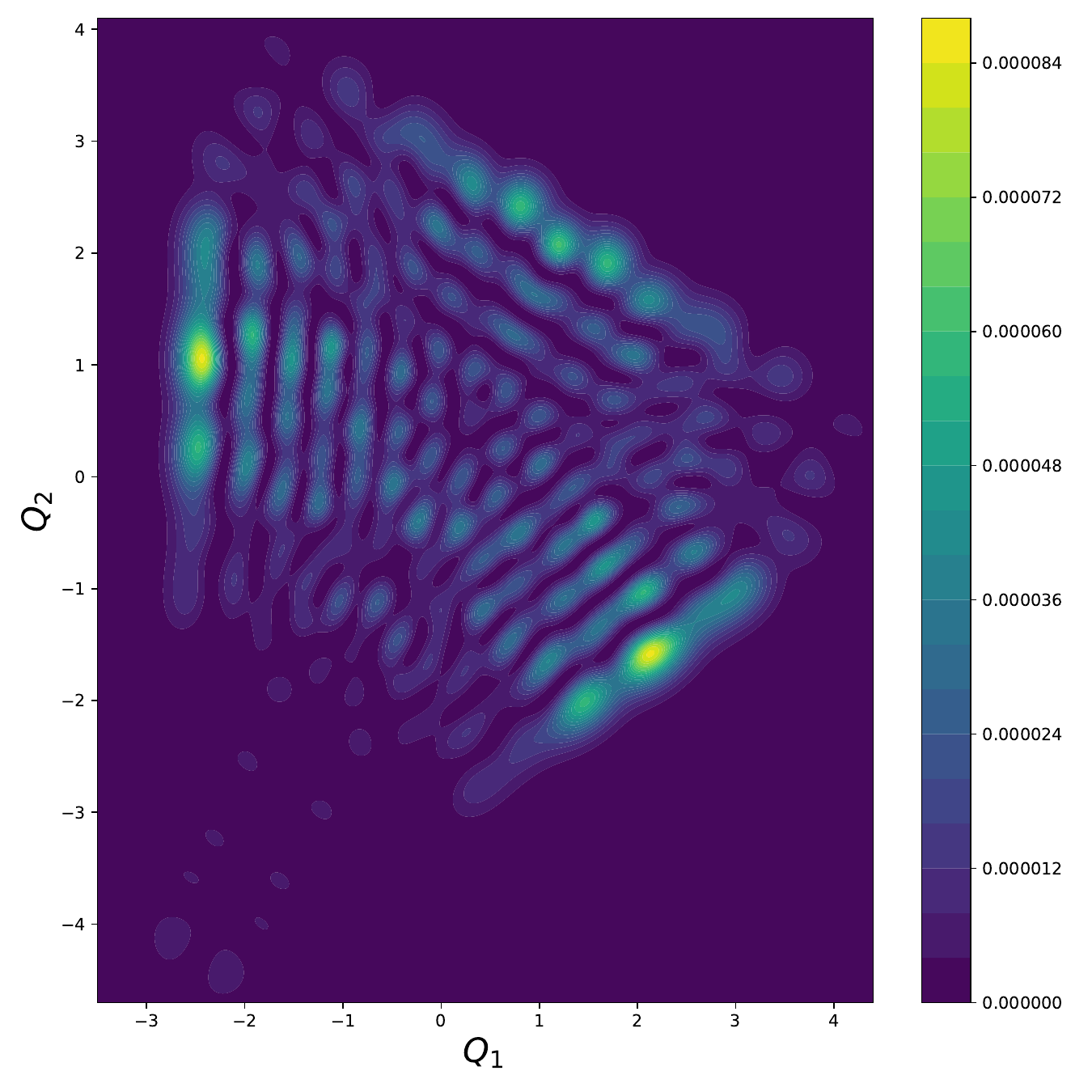}\quad
                \caption{}
                \label{ch_wf6}
        \end{subfigure}
	\medskip
        \begin{subfigure}{0.49\columnwidth}
                \centering
                \includegraphics[width=0.99\linewidth]{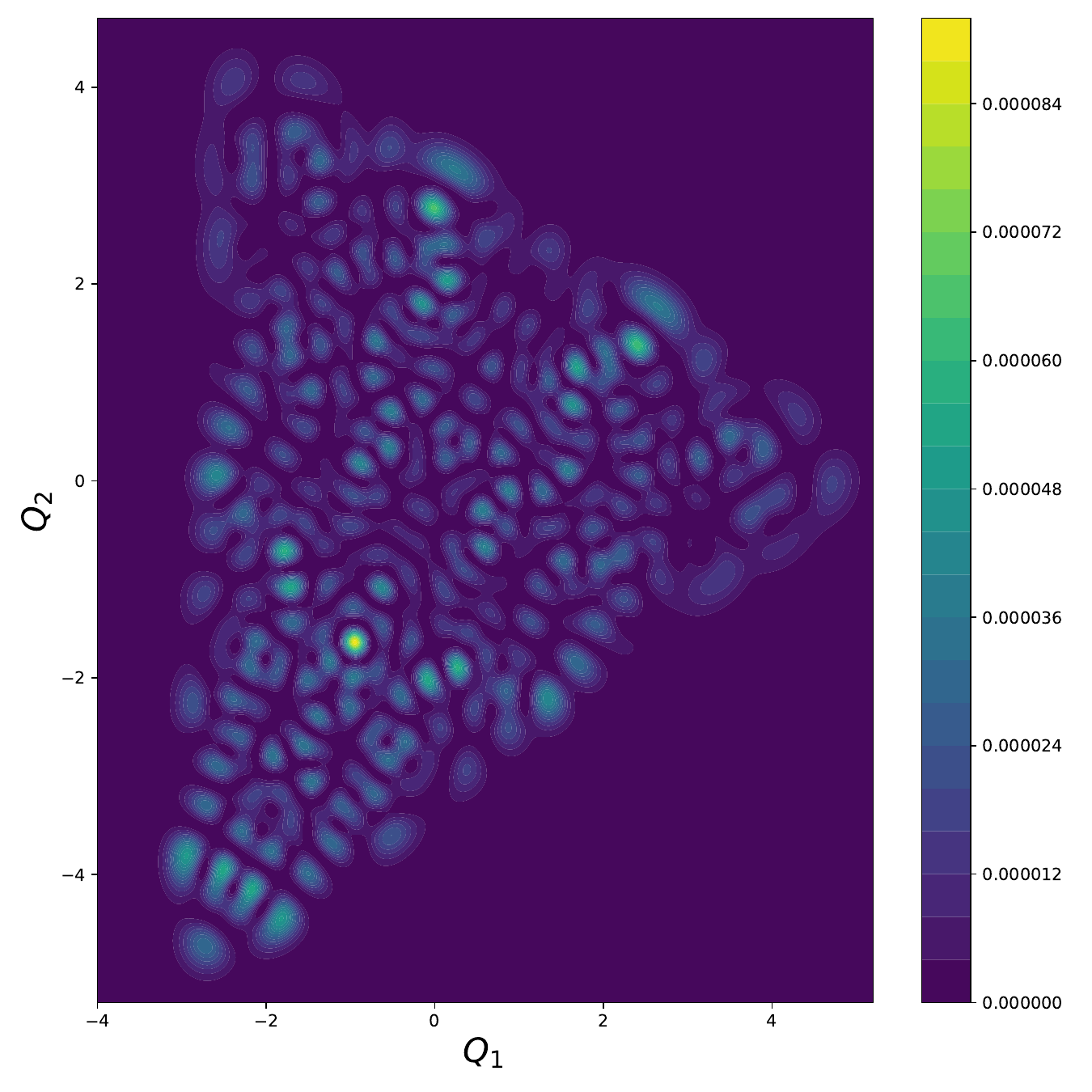}\quad
                \caption{}
                \label{ch_wf7}
        \end{subfigure}
        \begin{subfigure}{0.49\columnwidth}
                \centering
                \includegraphics[width=0.99\linewidth]{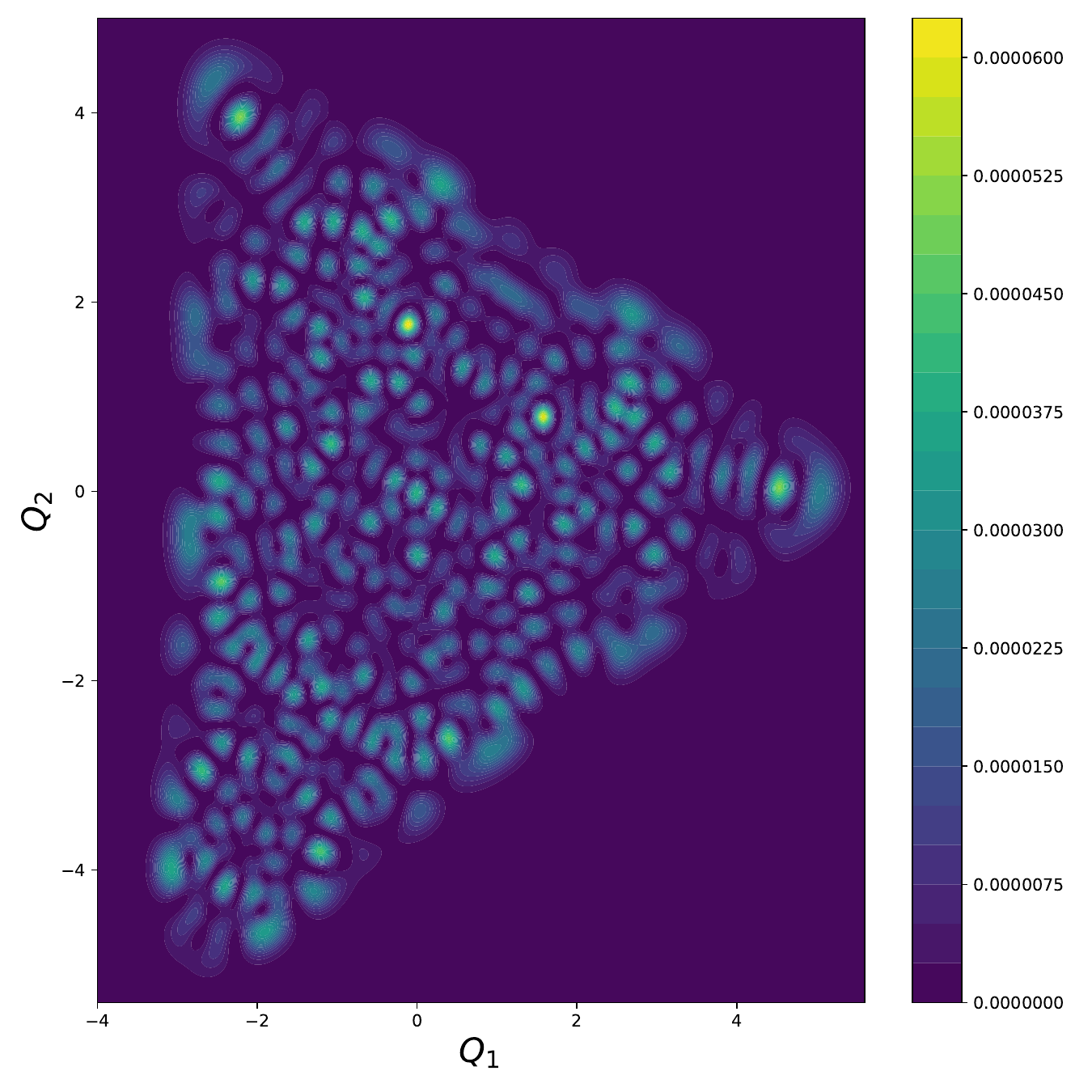}\quad
                \caption{}
                \label{ch_wf8}
        \end{subfigure}
	\caption{(Color online) Plot of $|\Phi|^{2}$ for $k=0$,$a=b=1$ and $\gamma = 1.6$. 
	(a) Ground state; (b) 1st excited state; (c) 2nd excited state; (d) 200th 
	excited state; (e) 300th excited state; (f) 400th excited state.}
	\label{wf-chaotic}
\end{figure}

The Bohr's correspondence principle allows to make a link between the classical physics and
the predictions of quantum mechanics for very high excited states. Thus, a chaotic  solution
in classical physics should bear some signature in the quantized version of the same system.
One of the objects to look for such signatures is the probability distribution of highly excited states.
The probability distribution of quantum states of  a classically integrable system is expected 
to be localized in space for low as well as very high quantum numbers. On the other hand,
the probability distribution of quantum states corresponding to a classically chaotic system
is expected to spread out in space for large values quantum number. The spreading out
of the probability distribution for highly excited states can be thought of as the quantum
analogue of the chaotic path taken in classical region. We plot the probability distribution
of the ground state and a few excited states in Fig. \ref{wf-integrable} and Fig. \ref{wf-chaotic}
for $\gamma=.2$ and $\gamma=1.6$, respectively. It is seen that the probability  distribution 
for $\gamma=0.2$ is localized with uniform distribution for low as well as highly excited states.
On the other hand, the complex behaviour arises in the highly excited wave-functions, and manifested
in the nonuniform probability distribution for $\gamma=1.6$.


\subsubsection{Level statistics \& Gap-ratio distribution}

The chaotic dynamics in classical physics is determined by studying the phase space trajectories,
Lyapunov exponents, Poincar\'e sections etc. The Schr\"odinger equation being a linear system, the
notion of quantum chaos is not uniform, and a quantification of the same is still an active area 
of research. One of the  standard approaches to explore signature of quantum chaos in a 
classically chaotic system is to study the level statistics. According to the BGS conjecture,
the quantum Hamiltonian with chaotic classical dynamics must fall into one of the three classical
ensembles of RMT\cite{bgs1984} \textemdash GOE, Gaussian unitary ensembles(GUE) and Gaussian
symplectic ensembles(GSE). The level statistics of quantum Hamiltonian for GOE with integrable
classical counterpart follow Poisson law  $P_P(s) = exp(-s)$, while it follows Wigner distribution
$P_{ W}(s) = \frac{\pi s}{2} exp\left( - \frac{\pi s^2}{4}\right)$ in the classical chaotic
region, where $s$ is the spacing of nearest-neighbour eigen-energies \cite{berry1977}.  

We study level statistics of the eigen spectrum of the Hamiltonian $H_{\textrm{eff}}$.
The level spacing distribution is shown by the probability density 
function $\rho(s)$ of the nearest neighbour spacing of unfolded eigen value. The procedure of 
unfolding the raw eigen value is a way of locally rescaling the energy spectra such that the mean
level density is one. The cumulative spectral function or the staircase function 
is defined as the number of levels with energy less than or equal to a 
certain value E. In particular, staircase function is defined as, $N(E) = \sum_{n} \Theta(E-E_n)$, 
where $\Theta$ is the unit step function. The function $N(E)$ consists of smooth part $N_{sm}(E)$ 
and a fluctuating part $N_{fl}(E)$, i. e. $N(E) = N_{sm} + N_{fl}$. We obtain $N_{sm}$ 
numerically, by fitting the staircase function with a polynomial of degree 15. 
The unfolded eigen values $x_{i}$ are obtained from raw eigen values $E_{i}$ as 
$x_i = N_{sm}(E_{i})$. Finally The level-spacing distributions are shown by the 
probability density function $\rho(s)$, where $s = x_{i+1} - x_{i}$. 

\begin{figure}[htbp]
	\begin{subfigure}{0.32\columnwidth}
		\centering
		\includegraphics[width=0.99\linewidth]{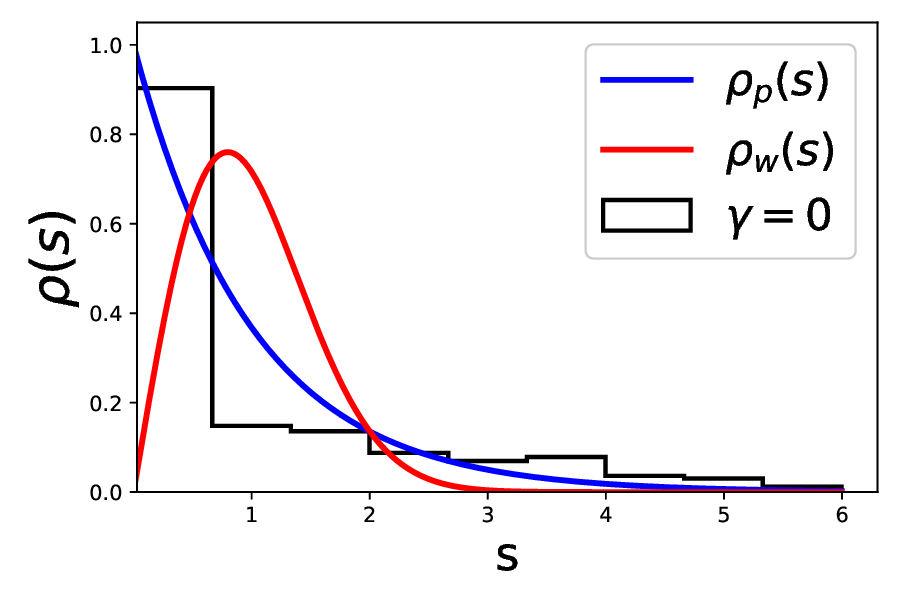}\quad
		\caption{}
		\label{fig:1}
	\end{subfigure}
	\begin{subfigure}{0.32\columnwidth}
		\centering
		\includegraphics[width=0.99\linewidth]{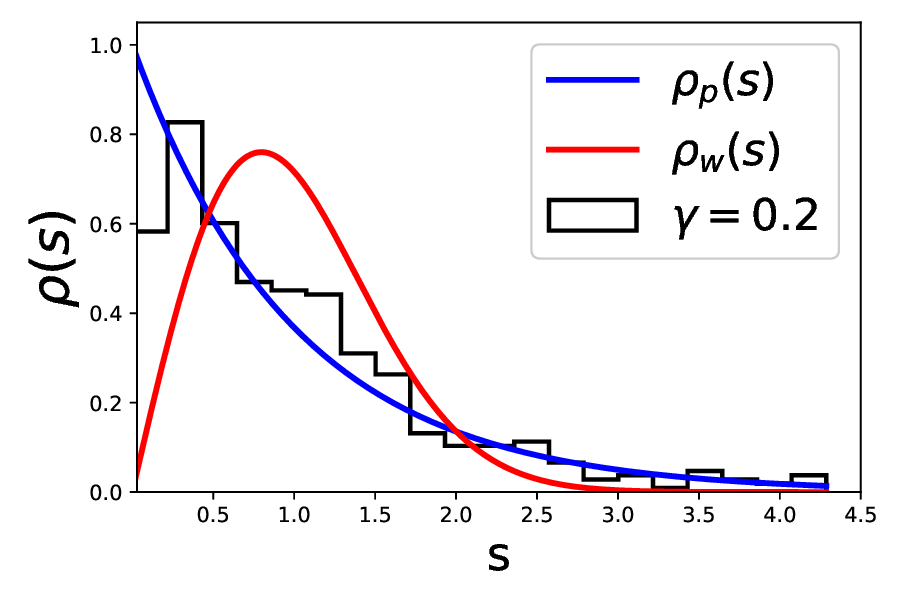}\quad
		\caption{}
		\label{fig:2}
	\end{subfigure}
	\begin{subfigure}{0.32\columnwidth}
                \centering
		\includegraphics[width=0.99\linewidth]{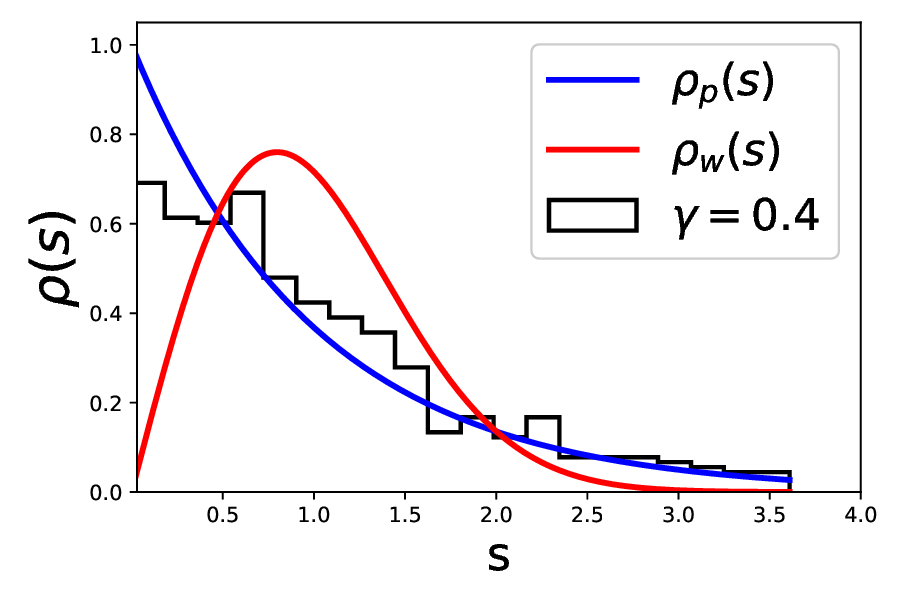}\quad
                \caption{}
                \label{fig:3}
        \end{subfigure}
	\medskip
	\begin{subfigure}{0.32\columnwidth}
                \centering
                \includegraphics[width=0.99\linewidth]{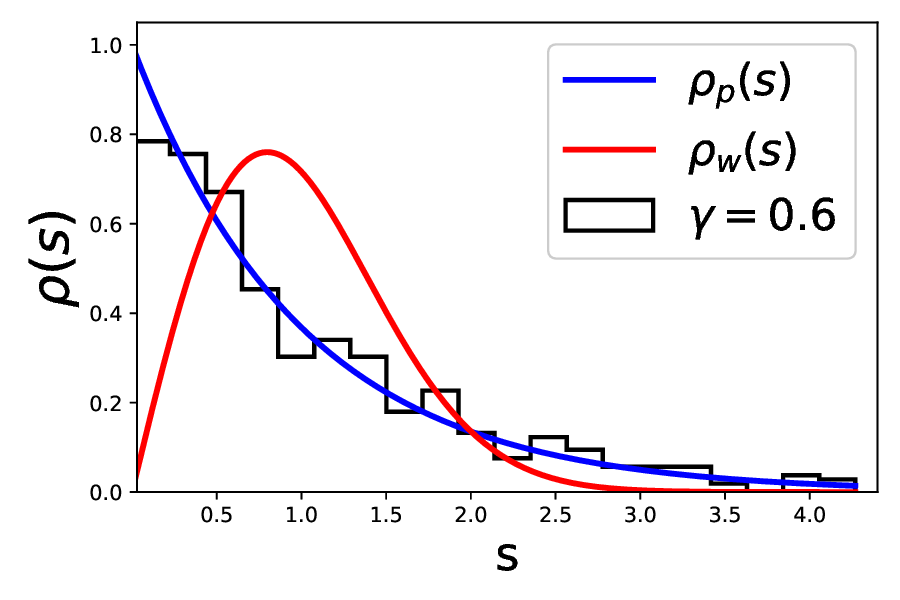}\quad
                \caption{}
                \label{fig:4}
        \end{subfigure}
	\begin{subfigure}{0.32\columnwidth}
                \centering
                \includegraphics[width=0.99\linewidth]{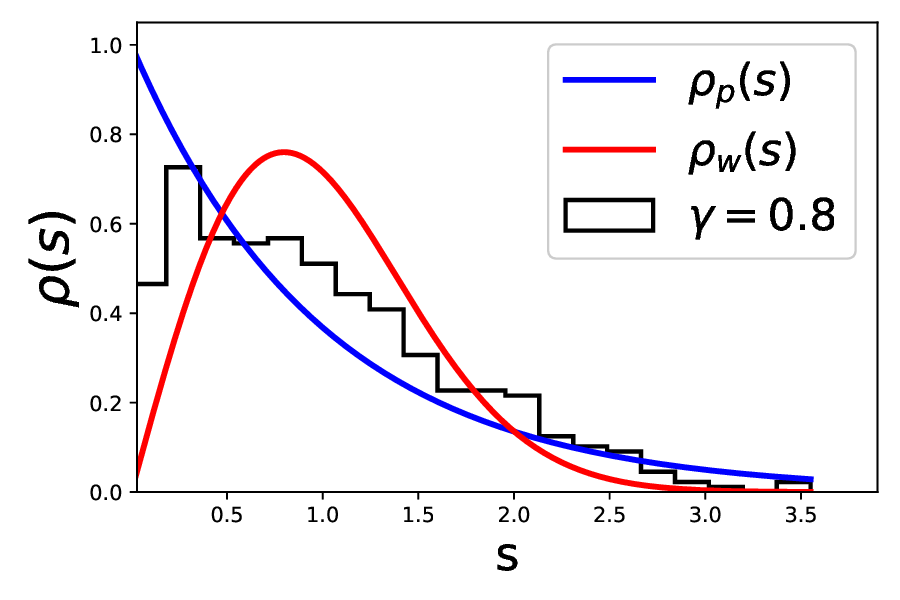}\quad
                \caption{}
                \label{fig:5}
        \end{subfigure}
	\begin{subfigure}{0.32\columnwidth}
                \centering
                \includegraphics[width=0.99\linewidth]{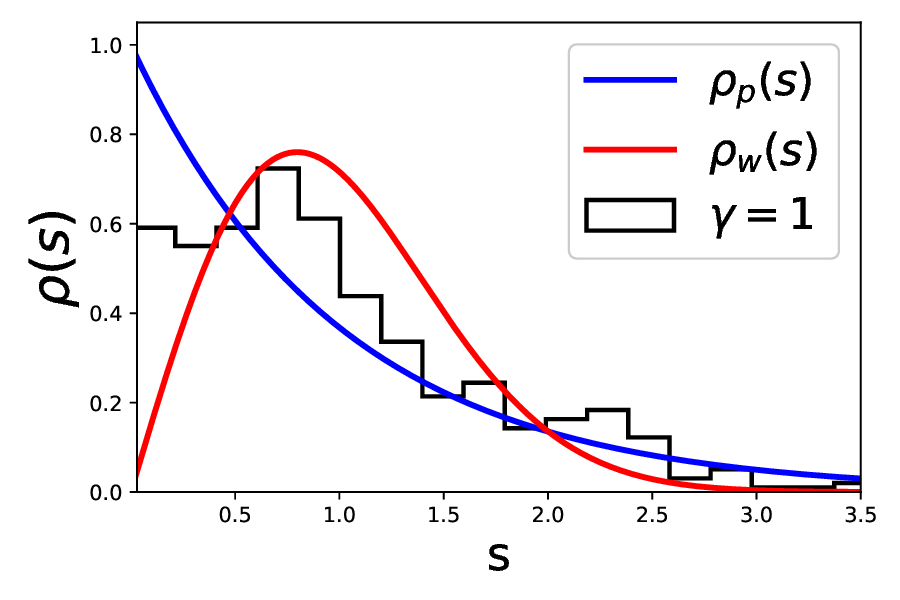}\quad
                \caption{}
                \label{fig:6}
        \end{subfigure}
	\medskip
	\begin{subfigure}{0.32\columnwidth}
                \centering
                \includegraphics[width=0.99\linewidth]{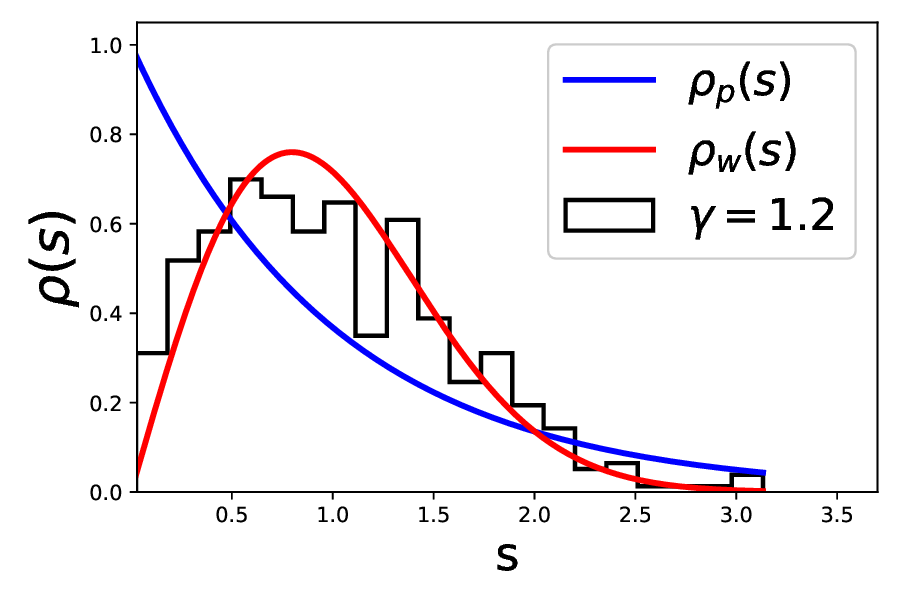}\quad
                \caption{}
                \label{fig:7}
        \end{subfigure}
	\begin{subfigure}{0.32\columnwidth}
                \centering
                \includegraphics[width=0.99\linewidth]{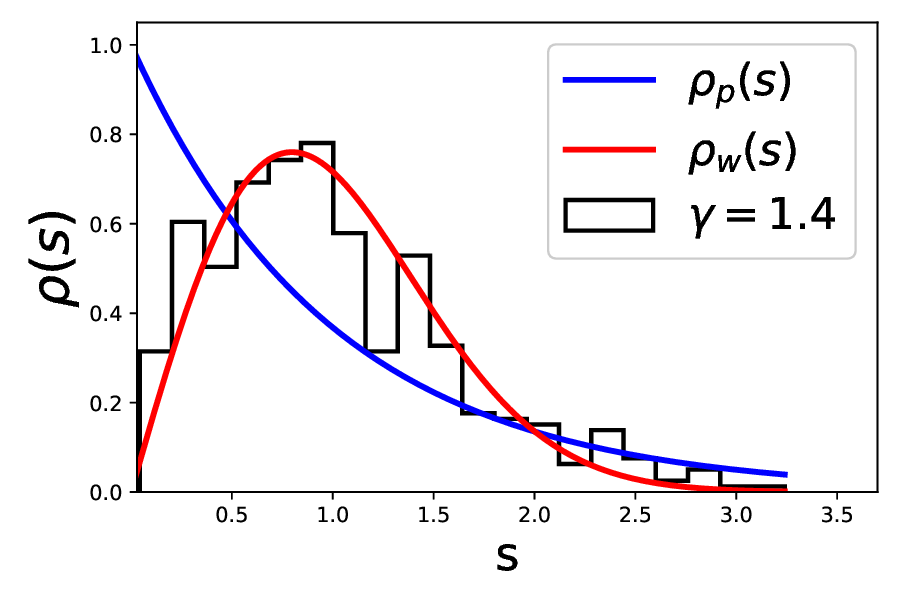}\quad
                \caption{}
                \label{fig:8}
	\end{subfigure}
	\begin{subfigure}{0.32\columnwidth}
                \centering
                \includegraphics[width=0.99\linewidth]{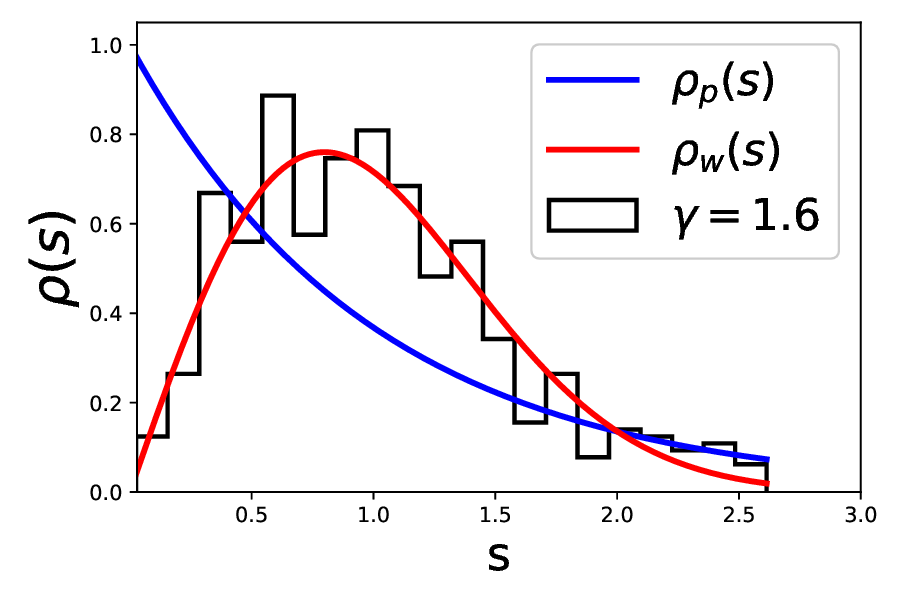}\quad
                \caption{}
                \label{fig:9}
        \end{subfigure}
	\medskip
        \begin{subfigure}{0.49\columnwidth}
                \centering
                \includegraphics[width=0.99\linewidth]{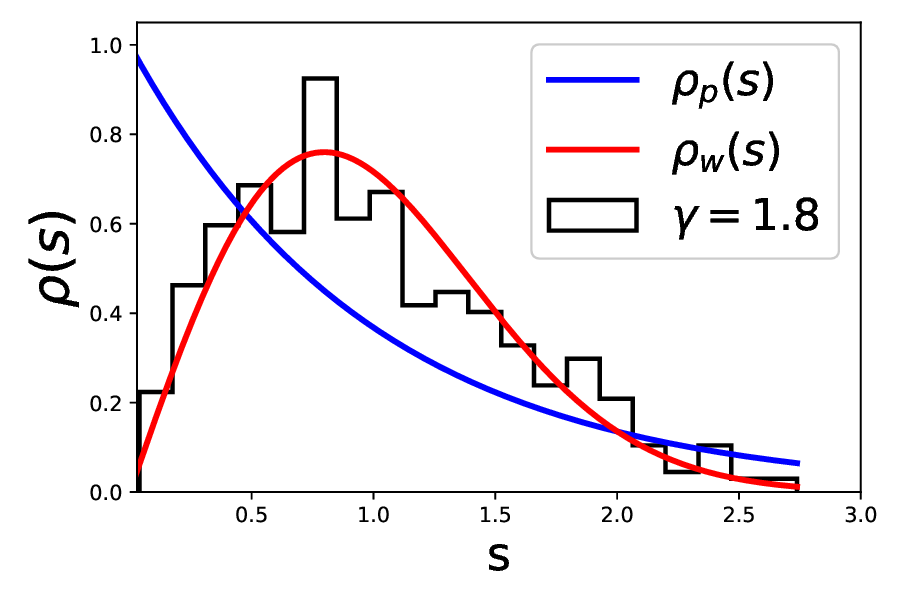}\quad
                \caption{}
                \label{fig:10}
        \end{subfigure}
        \begin{subfigure}{0.49\columnwidth}
                \centering
                \includegraphics[width=0.99\linewidth]{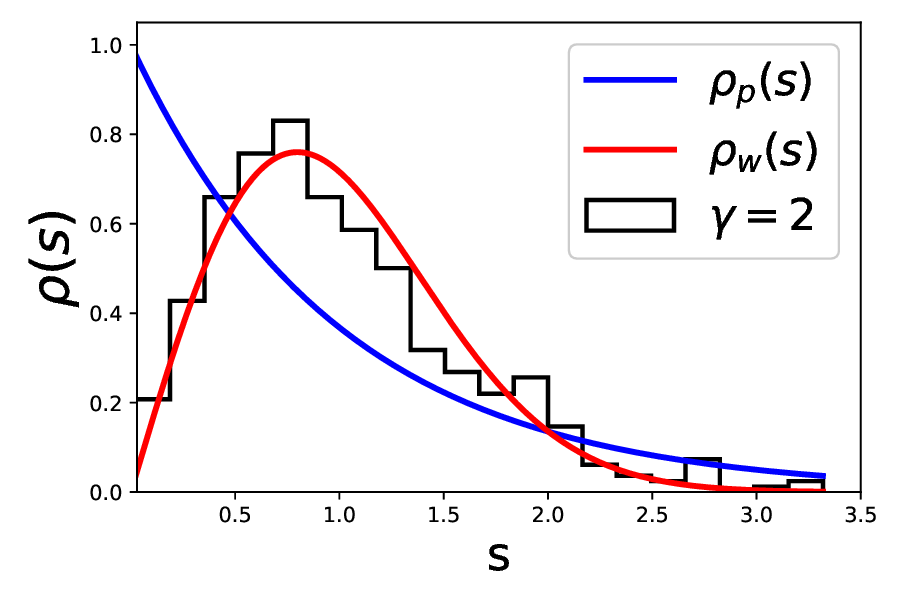}\quad
                \caption{}
                \label{fig:11}
        \end{subfigure}
	\caption{ (Color online) Plot of the level-spacing statistics of 
	Hamiltonian for different values of $\gamma$ and $k=0$, \ $a=b=1$.
	(a) $\gamma = 0$; (b) $\gamma = 0.2$; (c) $\gamma = 0.4$; (d) $\gamma = 0.6$;
	(e) $\gamma = 0.8$, (f) $\gamma = 1.0$, (g) $\gamma = 1.2$, (h) $\gamma = 1.4$,
	(i) $\gamma = 1.6$, (j) $\gamma = 1.8$, (k) $\gamma = 2.0$.}
	\label{level-spac}
\end{figure}
\begin{figure}[htbp]
        \begin{subfigure}{0.32\columnwidth}
                \centering
                \includegraphics[width=0.99\linewidth]{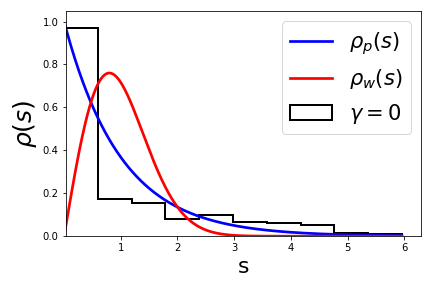}\quad
                \caption{}
                \label{ls_k1_g00}
        \end{subfigure}
        \begin{subfigure}{0.32\columnwidth}
                \centering
                \includegraphics[width=0.99\linewidth]{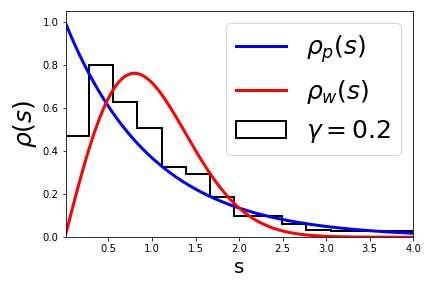}\quad
                \caption{}
                \label{ls_k1_g02}
        \end{subfigure}
        \begin{subfigure}{0.32\columnwidth}
                \centering
                \includegraphics[width=0.99\linewidth]{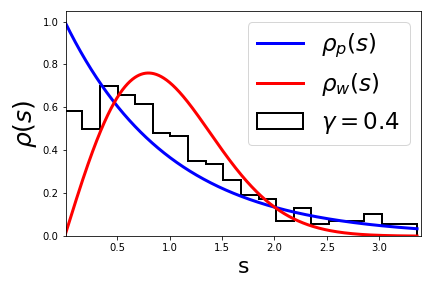}\quad
                \caption{}
                \label{ls_k1_g04}
        \end{subfigure}
        \medskip
        \begin{subfigure}{0.32\columnwidth}
                \centering
                \includegraphics[width=0.99\linewidth]{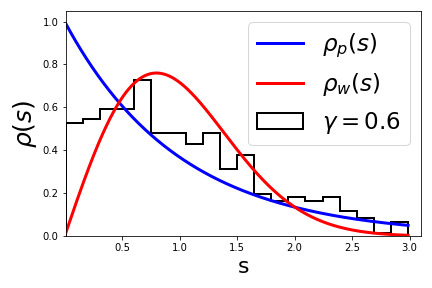}\quad
                \caption{}
                \label{ls_k1_g06}
        \end{subfigure}
        \begin{subfigure}{0.32\columnwidth}
                \centering
                \includegraphics[width=0.99\linewidth]{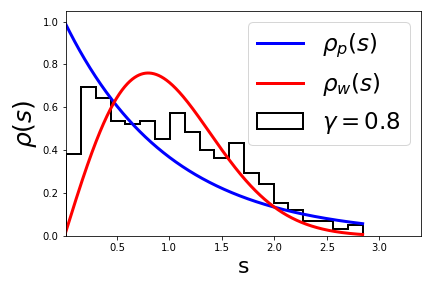}\quad
                \caption{}
                \label{ls_k1_g08}
        \end{subfigure}
        \begin{subfigure}{0.32\columnwidth}
                \centering
                \includegraphics[width=0.99\linewidth]{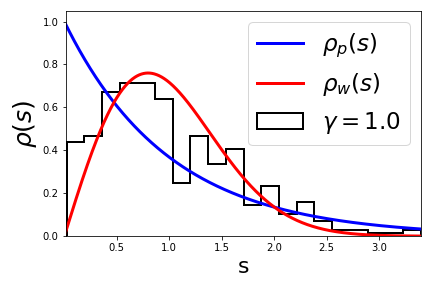}\quad
                \caption{}
                \label{ls_k1_g10}
        \end{subfigure}
        \medskip
        \begin{subfigure}{0.32\columnwidth}
                \centering
                \includegraphics[width=0.99\linewidth]{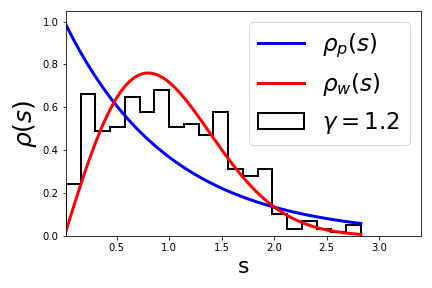}\quad
                \caption{}
                \label{ls_k1_g12}
        \end{subfigure}
        \begin{subfigure}{0.32\columnwidth}
		\centering
		\includegraphics[width=0.99\linewidth]{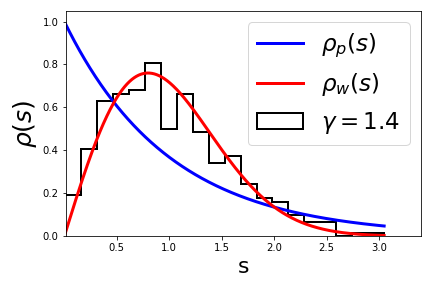}\quad
		\caption{}
		\label{ls_k1_g14}
	\end{subfigure}
	\begin{subfigure}{0.32\columnwidth}
                \centering
                \includegraphics[width=0.99\linewidth]{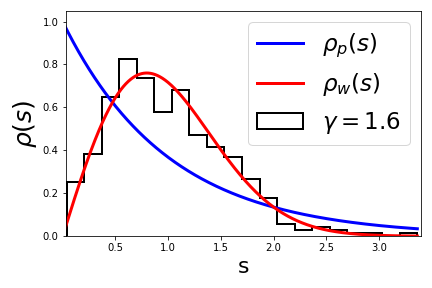}\quad
                \caption{}
                \label{ls_k1_g16}
        \end{subfigure}
	\medskip
	\begin{subfigure}{0.48\columnwidth}
                \centering
                \includegraphics[width=0.99\linewidth]{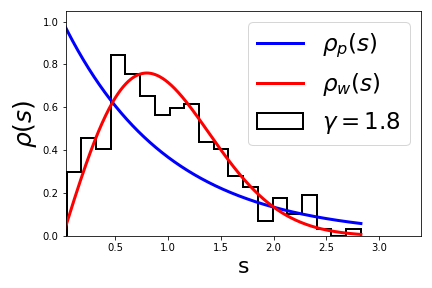}\quad
                \caption{}
                \label{ls_k1_g18}
        \end{subfigure}
        \begin{subfigure}{0.48\columnwidth}
                \centering
                \includegraphics[width=0.99\linewidth]{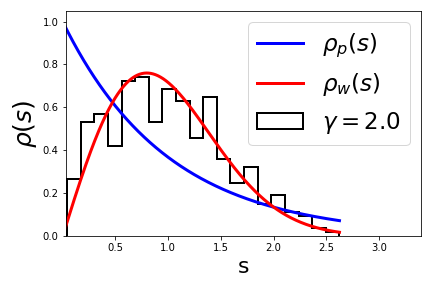}\quad
                \caption{}
                \label{ls_k1_g20}
        \end{subfigure}
	\caption{(Color online) Plot of the level-spacing statistics of
        Hamiltonian for different values of $\gamma$ and $k=1$ and $a=b=1$.
	(a) $\gamma = 0$; (b) $\gamma = 0.2$; (c) $\gamma = 0.4$; (d) $\gamma = 0.6$;
        (e) $\gamma = 0.8$, (f) $\gamma = 1.0$, (g) $\gamma = 1.2$, (h) $\gamma = 1.4$,
        (i) $\gamma = 1.6$, (j) $\gamma = 1.8$, (k) $\gamma = 2.0$.}
	\label{level-spac2}
\end{figure}

\begin{figure}[htbp]
	\begin{subfigure}{0.49\columnwidth}
	\centering
	\includegraphics[width = 0.98\linewidth]{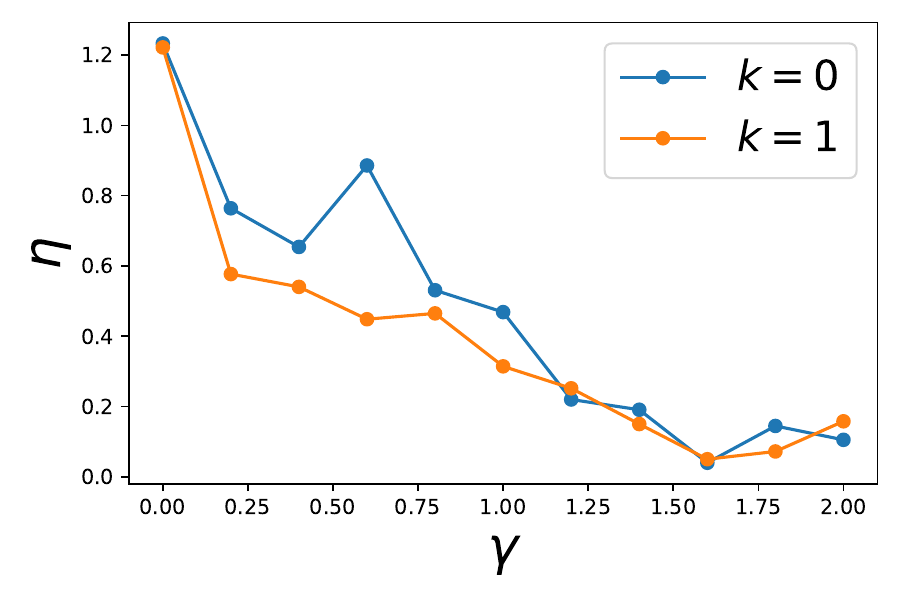}\quad
		\caption{}
		\label{bf1}
	\end{subfigure}
	\begin{subfigure}{0.49\columnwidth}
		\centering
		\includegraphics[width = 0.98\linewidth]{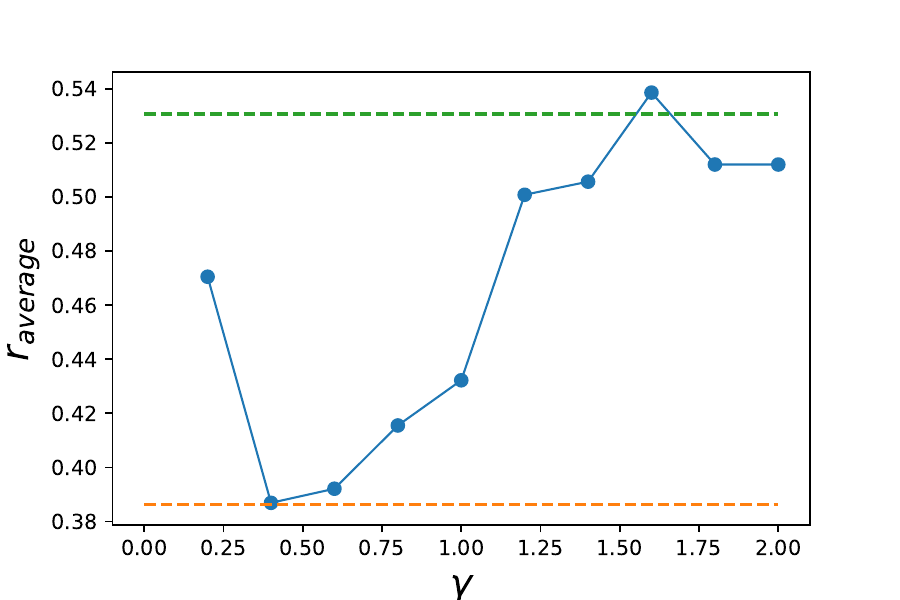}\quad
		\caption{}
		\label{bf2}
	\end{subfigure}
	\caption{(Color Online) (a) Plot of $\eta$ with respect to $\gamma$, for $k=0, \ 1$;
	(b) Plot of $\langle \tilde{r} \rangle$ with respect to $\gamma$ for $k=0$.}
	\label{bifur}
\end{figure}

The level spacing distributions of the set of eigen values of the Hamiltonian $H_{\textrm{eff}}$
for different values of $\gamma$ are shown in Fig. \ref{level-spac} and Fig. \ref{level-spac2}.
The blue and red solid lines represent the theoretical graphs for the Poisson and Wigner distributions,
respectively. It is seen by comparing Figs. \ref{level-spac} and \ref{level-spac2} that the nature of
the level spacing distributions are similar for $k=0$ and $k=1$. Further, $\rho(s)$ changes smoothly
from the Poisson to Wigner distribution via some intermediate distributions as $\gamma$ is varied from
$\gamma=0$ to $\gamma=2$. We define a quantity $\eta$ as follows,
\bea
\eta = \left\vert \frac{\int_{0}^{s_0} \left[ P(s) - P_W(s) \right] ds}{\int_{0}^{s_0} \left[ P_P(s) 
- P_W(s) \right] ds} \right\vert,
\eea
\noindent where $s_0 = 0.4729......$ is the intersection point of $P_P(s)$ and $P_W(s)$.
It may be noted that $\eta = 1$ for $P(s) = P_P(s)$, while $\eta = 0$ for $P(s) = P_W(s)$.
Thus, $\eta$ is an indicator of the nature of the distribution, and the plot of the $\eta$
with respect to the parameter $\gamma$  is shown in Fig. \ref{bf1}. This diagram in a sense
acts as the bifurcation diagram with respect to $\gamma$. 
The value of $\eta$ approaches one near $\gamma=0$ and goes to zero as $\gamma$ passes through
the value $\gamma \approx 1.5$ \textemdash the system transits from the integrable to the chaotic
region. 

The unfolding procedure during the calculation of level statistics is a bit cumbersome and 
sometime system dependent. The statistics of the ratio of two consecutive energy-gaps\cite{oganesyan}
has been proposed as an alternative measure for the same purpose. The spacing between adjacent energy
levels $\delta_{n} = E_{n+1} - E_{n}$, and the ratio between adjacent gaps is defined as,
\bea
0 \leq \tilde{r}_{n} = min\{\delta_{n},\delta_{n-1}\} / max\{\delta_{n},\delta_{n-1}\} \leq 1. \nonumber
\eea
\noindent The probability distribution of this ratio $\tilde{r}$ for integrable Hamiltonian, 
$P_P(\tilde{r}) = \frac{2}{(1+\tilde{r})^2}$ and the mean value of this is $\langle \tilde{r} \rangle_{P} 
= 2 \ ln 2 - 1 \cong 0.386$. The theoretical expression of probability distribution of $\tilde{r}$ 
in case of $GOE$ ensembles is, $P_{GOE}(\tilde{r}) = \frac{27}{4} \frac{\tilde{r} + \tilde{r}^2}
{(1 + \tilde{r} + \tilde{r}^2)^2}$. The mean value in this case, $\langle \tilde{r} \rangle_{GOE} 
= 0.5295 \pm 0.0006$. 

\begin{figure}[htbp]
        \begin{subfigure}{0.32\columnwidth}
                \centering
                \includegraphics[width=0.99\linewidth]{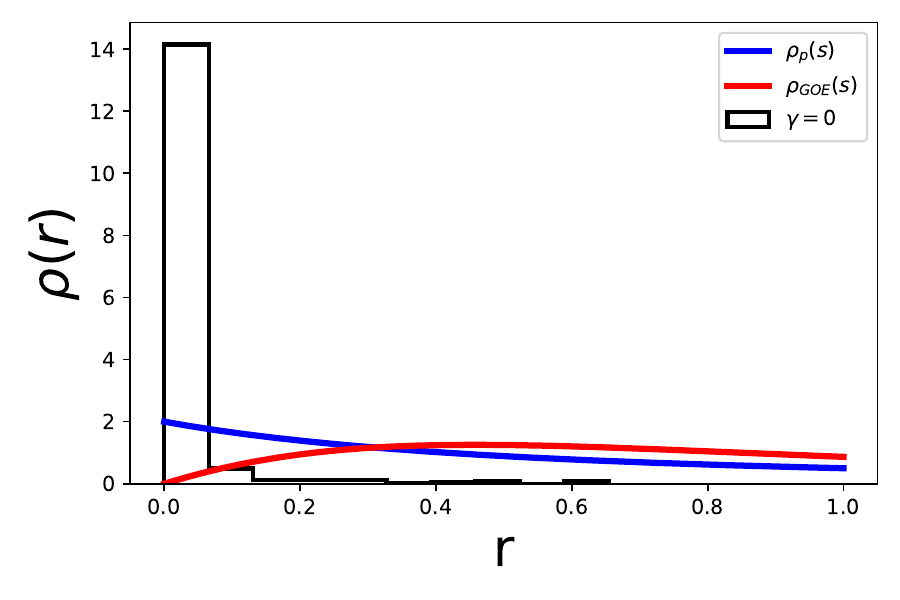}\quad
                \caption{}
                \label{gr:1}
        \end{subfigure}
        \begin{subfigure}{0.32\columnwidth}
                \centering
                \includegraphics[width=0.99\linewidth]{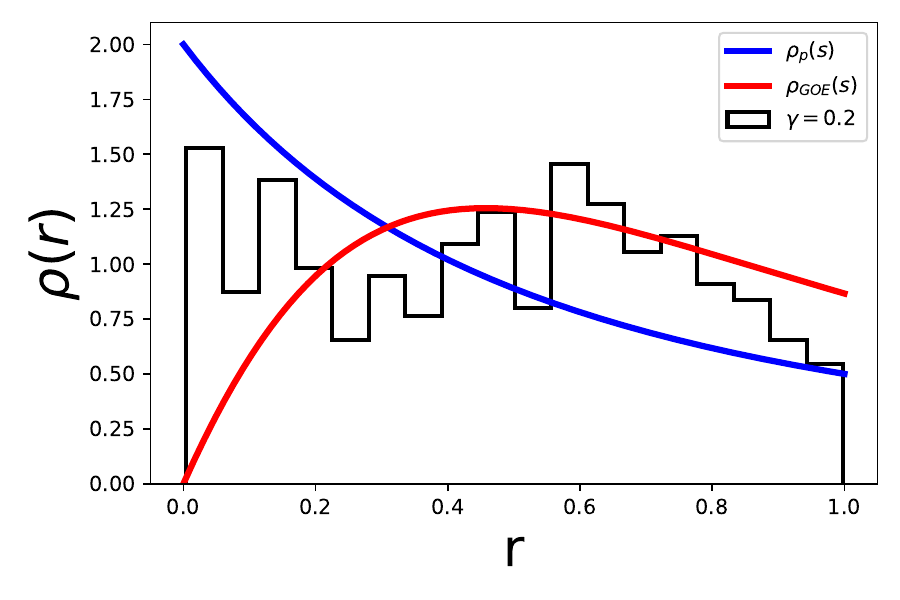}\quad
                \caption{}
                \label{gr:2}
        \end{subfigure}
        \begin{subfigure}{0.32\columnwidth}
                \centering
                \includegraphics[width=0.99\linewidth]{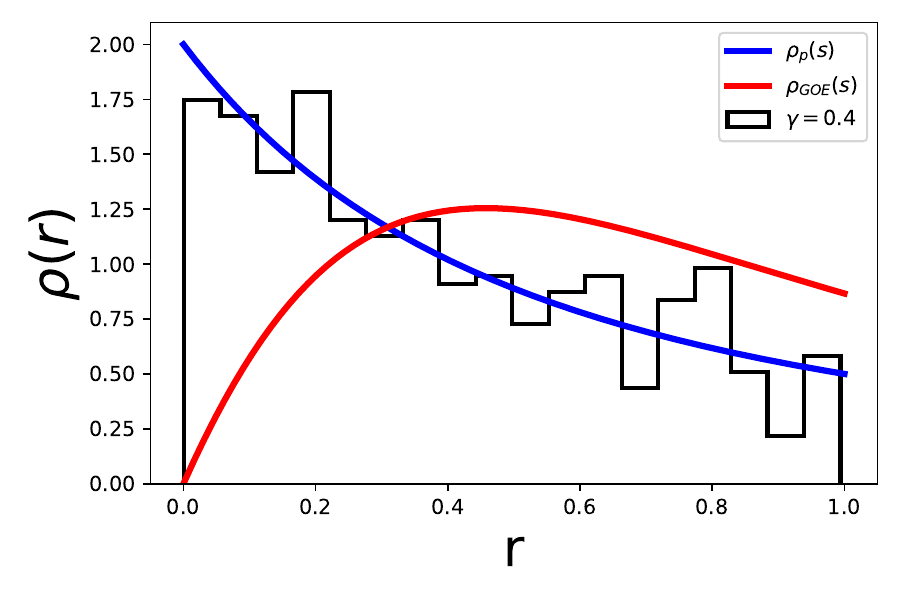}\quad
                \caption{}
                \label{gr:3}
        \end{subfigure}
        \medskip
        \begin{subfigure}{0.32\columnwidth}
                \centering
                \includegraphics[width=0.99\linewidth]{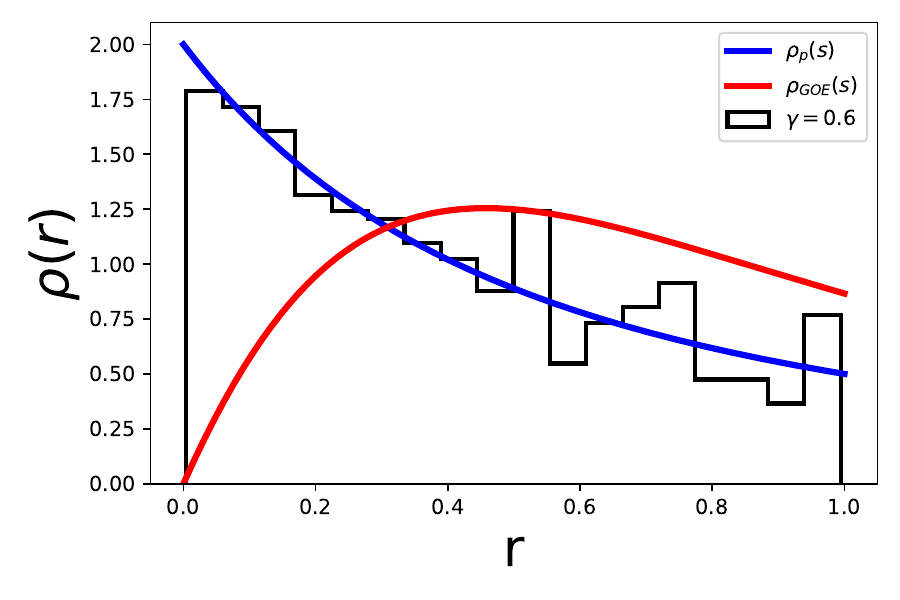}\quad
                \caption{}
                \label{gr:4}
        \end{subfigure}
        \begin{subfigure}{0.32\columnwidth}
                \centering
                \includegraphics[width=0.99\linewidth]{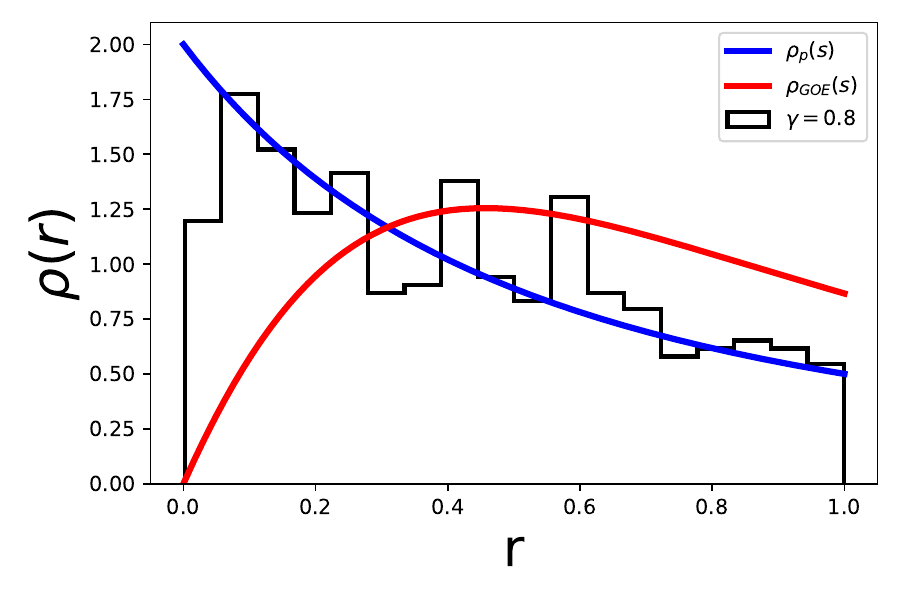}\quad
                \caption{}
                \label{gr:5}
        \end{subfigure}
        \begin{subfigure}{0.32\columnwidth}
                \centering
                \includegraphics[width=0.99\linewidth]{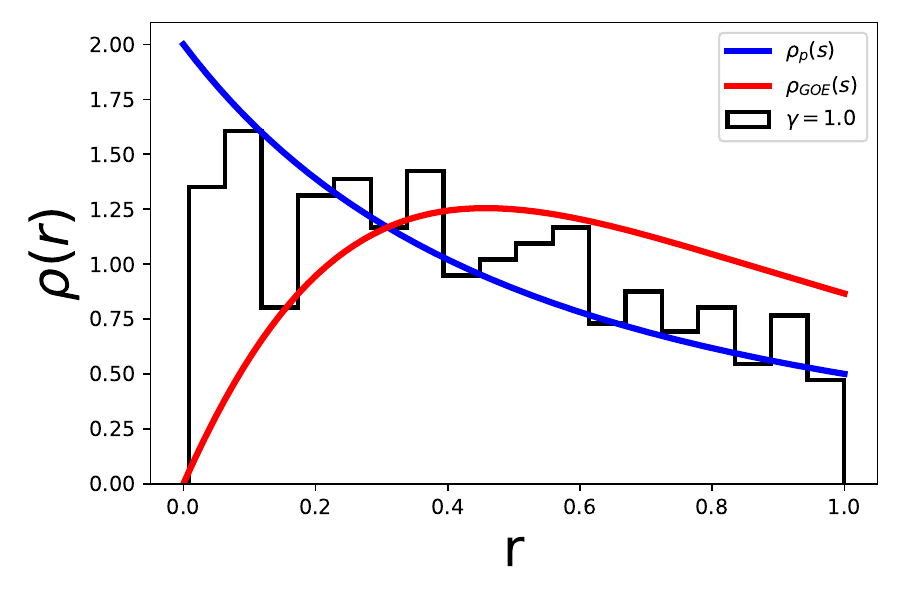}\quad
                \caption{}
                \label{gr:6}
        \end{subfigure}
        \medskip
        \begin{subfigure}{0.32\columnwidth}
                \centering
                \includegraphics[width=0.99\linewidth]{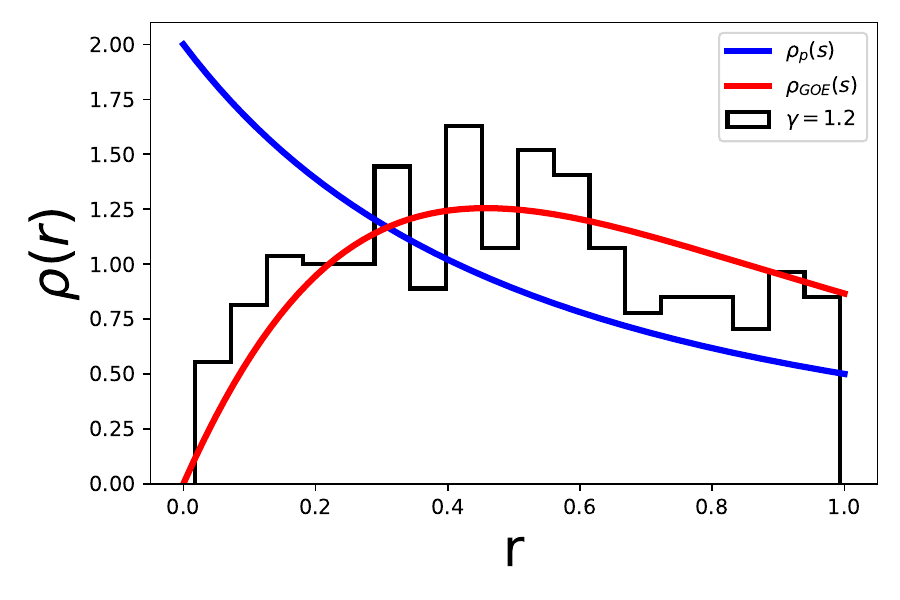}\quad
                \caption{}
                \label{gr:7}
        \end{subfigure}
	\begin{subfigure}{0.32\columnwidth}
                \centering
                \includegraphics[width=0.99\linewidth]{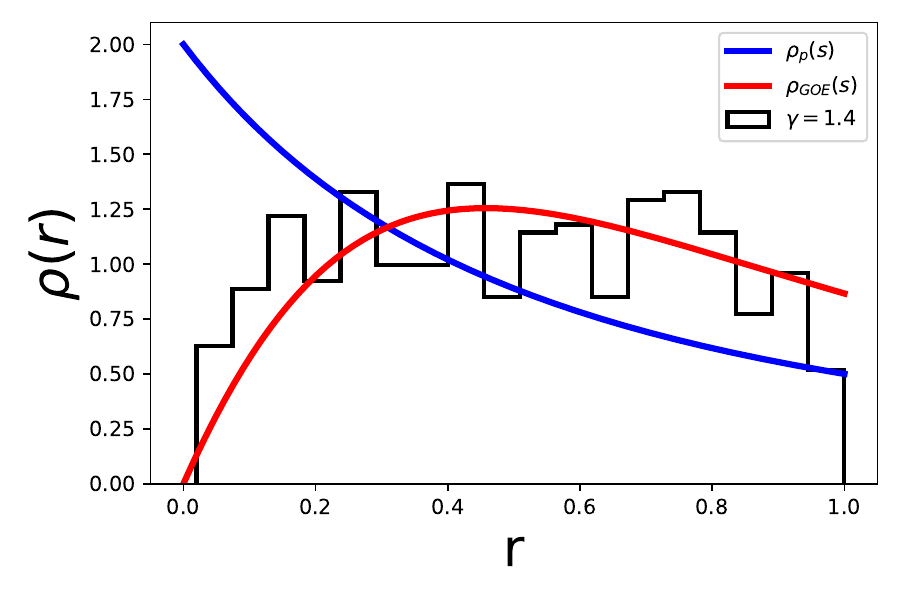}\quad
                \caption{}
                \label{gr:5}
        \end{subfigure}
        \begin{subfigure}{0.32\columnwidth}
                \centering
                \includegraphics[width=0.99\linewidth]{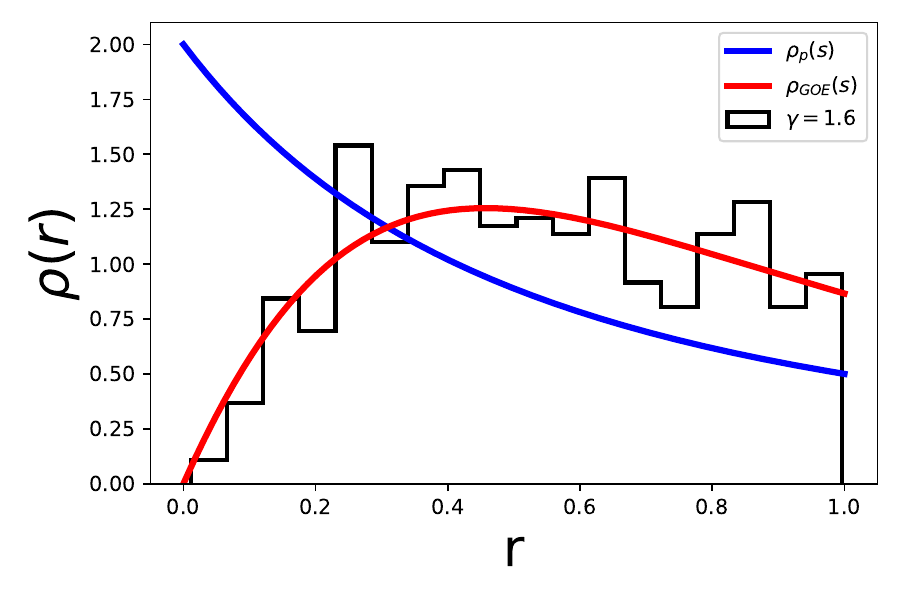}\quad
                \caption{}
                \label{gr:6}
        \end{subfigure}
	\medskip
        \begin{subfigure}{0.49\columnwidth}
                \centering
                \includegraphics[width=0.99\linewidth]{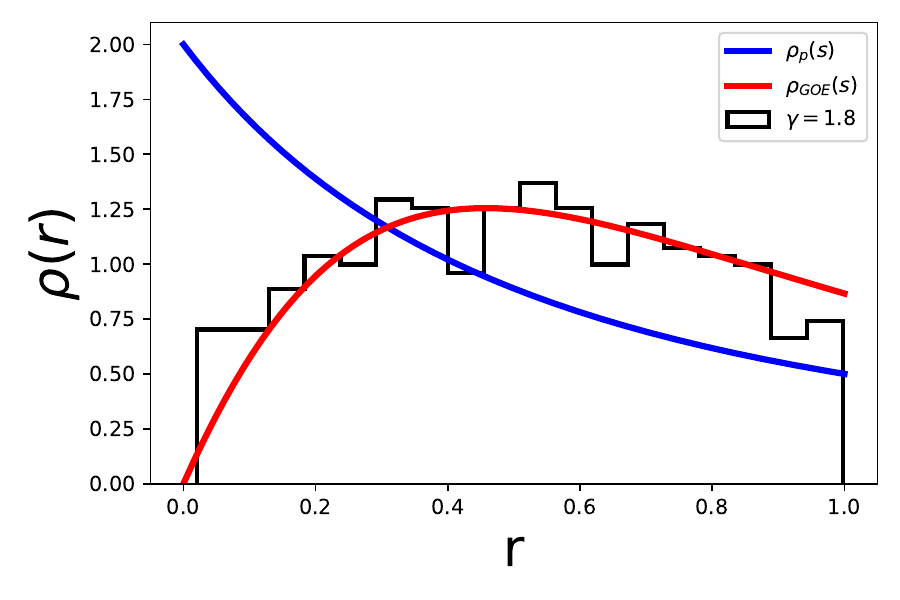}\quad
                \caption{}
                \label{gr:10}
        \end{subfigure}
        \begin{subfigure}{0.49\columnwidth}
                \centering
                \includegraphics[width=0.99\linewidth]{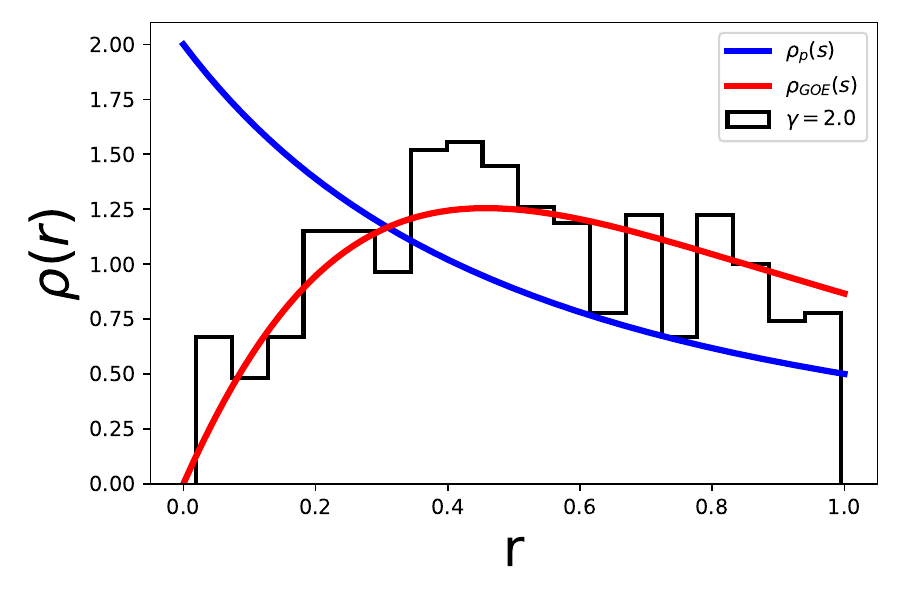}\quad
                \caption{}
                \label{gr:11}
        \end{subfigure}
	\caption{ (Color online) Plot of the Gap Ration $\tilde{r}$  statistics of
        Hamiltonian with $k=0$,$a=b=1$ for different values of $\gamma$.
	(a) $\gamma = 0$; (b) $\gamma = 0.2$; (c) $\gamma = 0.4$; (d) $\gamma = 0.6$;
        (e) $\gamma = 0.8$, (f) $\gamma = 1.0$, (g) $\gamma = 1.2$, (h) $\gamma = 1.4$,
        (i) $\gamma = 1.6$, (j) $\gamma = 1.8$, (k) $\gamma = 2.0$.}
        \label{gapratio}
\end{figure}
\noindent We study the gap ratio distribution for $k=0$ only because we have already shown that 
the change in $k$ does not alter the nature of level spacing distribution. 
The Fig. \ref{gapratio} shows that the probability density distribution 
of gap ratio $\tilde{r}$ is close to the Poisson distribution up to  $\gamma \approx 1.0$. 
This signifies integrable system. When $\gamma \approx 1.4$ the distribution goes close to 
the theoretical value of the probability density distribution of GOE ensembles. 
This signifies the non-integrable region. The Fig. (\ref{bf2}) is the bifurcation diagram for 
this gap ratio distribution. It is see from the bifurcation diagram that while $\gamma \gtrapprox 1.4$ 
the average value of gap ratio is close to 0.52 i.e. the system is chaotic. We show the 
quantum transition from integrable to chaotic region by studying nearest neighbour level 
spacing and gap ratio distribution. 

\section{Conclusions \& Discussions}

We have studied the periodic quantum Toda lattice with BLG for two and
three particles. The two-particle Toda lattice is integrable and we have constructed
two integrals of motion which are in involution. The translation invariance of the
system has been used to separate out the center of mass motion, and the effective Hamiltonian
in the relative coordinate is described by a particle moving in a potential consisting
of harmonic plus cosine hyperbolic. The effective potential describes
single or double wells depending on the strength of the loss-gain and the exponential
terms. The eigen value equation has been solved numerically, and eigen energies as well as
eigenfuntion corresponding to quantum bound states for a few low lying states have been
presented. The qualitative behaviour is similar to that of single of double well arising 
out of a Quartic potential.

The quantum Toda lattice with BLG and VMC for three particles is translation invariant.
The effective Hamiltonian after the separation of the center of mass may be interpreted
as a particle moving in a two dimensional potential, consisting of anisotropic harmonic oscillators
plus Toda potential, and subjected to external uniform magnetic field with its strength being
linearly proportional to the strength of the BLG. The angular frequencies of the  harmonic oscillators
depend quadratically on the strength of the BLG. It has been shown that the effective Hamiltonian
is exactly solvable in the limit in which the Toda potential reduces to coupled oscillators.
The eigenspectra and eigenfunctions have been obtained analytically by mapping the
effective Hamiltonian to that of decoupled anisotropic oscillators in two dimensions
via a similarity transformation. There is a limit in which reduction of a degree of freedom occur 
in phase space and spectra becomes infinitely degenerate.

We have obtained two integrals of motion which are in involution. Although we have not
found the third integral of motion which is required to show the complete integrability,
the numerical investigations indicates that the three-particle quantum Toda system is integrable
below a critical value of the BLG strength and chaotic above this critical value. The
quantum integrability and chaos have been investigated via level statistics as well as
gap-ratio distribution of the level-spacing of nearest-neighbour eigen values. In particular,
we have observed the quantum transition from the chaotic to the integrable region,
when the loss-gain strength crosses a critical value and goes to zero  \textemdash
the level spacing as well as the gap-ratio distributions smoothly changes from the Wigner-Dyson
distribution and tends to follow the Poisson distribution. We have also studied the level
repulsion phenomena in the energy spectra in both the two and three particle Toda lattices.
It has been shown through the graphical presentations that the degree of level repulsion is
large in the case of the three particle system. The quantum transition from chaotic to
integrable region has also been independently confirmed by studying the complexity
in higher order excited state wave functions.

There are some immediate questions which may be pursued for future investigations. For
example, the analytic expression for the third integral of motion is yet unknown. The
standard technique of LAX pair formalism or Painelev\'e analysis may be employed
to find an analytic expression for the same. Further,
the generic problem for $N>3$ particles is worth pursuing to see whether or not the
mixed phases of intergrability and chaos persist for arbitrary number of particles.
It is possible that the system may not be integrable for any range of the strength
of the BLG for large $N$. Finally, a semi-classical approach to understand the
integrability and chaos is desirable. Some of these issues will be addressed 
in future publications.

\section{Acknowledgements}

The work of SG is supported by Inspire fellowship(Inspire Code: IF190276) of Govt. of India


\begin{thebibliography}{10}
	\bibitem{toda1} M. Toda, Vibration of a Chain with Nonlinear Interaction, Journal of the 
		Physical Society of Japan {\bf 22(2)}, 431, 1967.
	\bibitem{toda2} M. Toda, Studies of a non-linear lattice, Phys. Rep. {\bf 18}, 1 (1975).
	\bibitem{toda3} M. Toda, Theory of Nonlinear Lattices, 2nd enl. ed., Springer, Berlin, 1989.
	\bibitem{date1976} E. Date and S. Tanaka, Periodic multi-soliton solutions of Korteweg-de Vries equation and
		Toda lattice, Prog. Theor. Phys. {\bf 59}, 107 (1976).
	\bibitem{flaschka} H. Flaschka, The Toda lattice. II. Existence of integrals, Phys. Rev. B {\bf 9}, 1924 (1974).
	\bibitem{henon} M. H\'enon, Integrals of the Toda lattice, Phys. Rev. B {\bf 9}, 1921 (1974). 
	\bibitem{agrotis} M. A. Agrotis, P. A. Damianou and C. Sophocleous, The Toda lattice is super-integrable,
		Physica A {\bf 365}, 235 (2006).
	\bibitem{toda4} M. Toda, Solitons and Heat Conduction, Phys. Scr. {\bf 20}, 424 (1979).
	\bibitem{sataric} M. Sataric, J. A. Tuszynski, R. Zakula, and S. Zekovic, Heat conductivity of a perturbed
		monatomic Toda lattice without impurities, J. Phys.: Condens. Matter {\bf 6}, 3917 (1994).
	\bibitem{muto} V. Muto, A. C. Scott, and P. L. Christiansen, A Toda lattice model for DNA: Thermally
		generated solitons, Physica D {\bf 44}, 75 (1990).
	\bibitem{dovivio} F. $d'$Ovivio, H. G. Bohr, and P. Lindgard, Solitons on H Bond in Proteins, J. Phys.: Con-
		dens. Matter {\bf 15}, S1699 (2003).
	\bibitem{oppo} G. L. Oppo and A. Politi, Toda potential in laser equations. Z. Phys. B-Condensed Matter
		{\bf 59}, 111 (1985).
	\bibitem{lien} Y. Lien, S. M. de Vries, M. P. van Exter, and J. P. Woerdman, Lasers as Toda oscillators,
		J. Optical Soc. Am. B {\bf 19(6)}, 1461–1466 (2002).
	\bibitem{cialdi} S. Cialdi, F. Castelli, and F. Prati, Lasers as Toda oscillators: An experimental 
		confirmation, Optics Communications {\bf 287}, 176 (2013).
	\bibitem{habib} S. Habib, H. E. Kandrup, and M. E. Mahon, Chaos and noise in a truncated Toda potential,
		Phys. Rev. E {\bf 53}, 5473 (1996).
	\bibitem{casati} G. Casati and J. Ford, Stochastic transition in the unequal-mass Toda lattice, Phys. Rev. 
		A {\bf 12}, 1702 (1975).
	\bibitem{vergara} L. Vergara and B. A. Malomed, Suppression of the generation of defect modes by a moving
		soliton in an inhomogeneous Toda lattice, Phys. Rev. E {\bf 77}, 047601 (2008).
	\bibitem{ezawa} M. Ezawa, Topological Edge States and Bulk-edge Correspondence in Dimerized Toda Lattice, 
		J. Phys. Soc. Jpn. {\bf 91}, 024703 (2022).
	\bibitem{proy} P. Roy and P. K. Ghosh, Balanced loss-gain induced chaos in a periodic Toda lattice,
 Phys. Lett. A {\bf 489}, 129156(2023).

	\bibitem{pkg} P. K. Ghosh, Classical Hamiltonian Systems with balanced loss and gain, J. Phys.: Conf.
                Ser. {\bf 2038}, 012012 (2021).
	\bibitem{bender} C. M. Bender, M. Gianfreda, S. K. Ozdemir, B. Peng, and L. Yang, Twofold transition in
		PT-symmetric coupled oscillators, Phys. Rev. A {\bf 88}, 062111 (2013).
	\bibitem{bender2} C. M. Bender, M. Gianfreda and S. P. Klevansky, Systems of coupled PT-symmetric oscil-
		lators, Phys. Rev A {\bf 90}, 022114 (2014).
	\bibitem{cuevas} Jes\'us Cuevas, Panayotis G. Kevrekidis, Avadh Saxena, and Avinash Khare, PT-symmetric
		dimer of coupled nonlinear oscillators, Phys. Rev. A {\bf 88}, 032108 (2013).

\bibitem{pkg2} P. K. Ghosh, Constructing solvable models of vector non-linear Schr\"odinger equation with balanced
loss and gain via non-unitary transformation, Phys. Lett. A {\bf 402}, 127361(2021).
	\bibitem{sg1} S. Ghosh and P. K. Ghosh, Non-linear Schr\"odinger equation with time-dependent balanced loss-gain 
		and space–time modulated non-linear interaction, Annals of Physics {\bf 454}, 169330 (2023).
	\bibitem{sg2} S. Ghosh and P. K. Ghosh, Solvable limits of a class of generalized vector nonlocal nonlinear 
		Schr\"odinger equation with balanced loss-gain, Phys. Scr. {\bf 98}, 115214(2023).

	\bibitem{barashenkov} I. V. Barashenkov and M. Gianfreda, An exactly solvable PT -symmetric dimer from a
Hamiltonian system of nonlinear oscillators with gain and loss, J. Phys. A: Math. Theor. {\bf 47}, 282001 (2014).
	\bibitem{ds1} D. Sinha and P. K. Ghosh, ${\cal{PT}}$-symmetric rational Calogero model with balanced loss and
		gain, Eur. Phys. J. Plus {\bf 132}, 460 (2017). 
	\bibitem{khare1} A. Khare and A. Saxena, Integrable oscillator type and Schrödinger type dimers, J. Phys.
		A: Math. Theor. {\bf 50}, 055202 (2017).
	\bibitem{ds2} P. K. Ghosh and Debdeep Sinha, Hamiltonian formulation of systems with balanced loss-
		gain and exactly solvable models, Annals of Physics {\bf 388}, 276 (2018).
	\bibitem{ds3} D. Sinha and P. K. Ghosh, On the bound states and correlation functions of a class of
Calogero-type quantum many-body problems with balanced loss and gain, J. Phys. A: Math.
		Theor. {\bf 52}, 505203 (2019).
	\bibitem{ds4} D. Sinha and P. K. Ghosh, Integrable coupled Li\'enard-type systems with balanced loss and
		gain, Annals of Physics {\bf 400}, 109 (2019).
	\bibitem{proy1} P. K. Ghosh and P. Roy, On regular and chaotic dynamics of a non-$\mathcal{PT}$-symmetric
Hamiltonian system of a coupled Duffing oscillator with balanced loss and gain, J. Phys. A: Math. Theor.
                {\bf 53}, 475202 (2020).
        \bibitem{proy2} P. Roy and P. K. Ghosh, Complex dynamical properties of coupled Van der Pol–Duffing oscillators with
                balanced loss and gain, J. Phys. A : Math. theor. {\bf 55}, 315701 (2022).
	\bibitem{pkg1} Pijush K. Ghosh, Taming Hamiltonian systems with balanced loss and gain via Lorentz
interaction : General results and a case study with Landau Hamiltonian, J. Phys. A: Math.
		Theor. {\bf 52}, 415202 (2019). 
	\bibitem{haake} F. Haake, Quantum Signatures of Chaos (Springer-Verlag, Berlin, 1991)
        \bibitem{bgs1984} O. Bohigas, M. J. Giannoni, and C. Schmit, Characterization of
                Chaotic Quantum Spectra and Universality of Level Fluctuation Laws, Phys. Rev.
                Lett. {\bf 52}, 1 (1984).
        \bibitem{berry1977} M. V. Berry and M. Tabor, Level clustering in the regular spectrum,
                Proc. R. Soc. Lond. A. {\bf 356}, 375 (1977).
        \bibitem{montambaux} G. Montambaux, D. Poilblanc, J. Bellissard, and C. Sire, Quantum Chaos
                in Spin-Fermion Models, Phys. Rev. Lett. {\bf 70}, 497 (1993).
        \bibitem{poilblanc} D. Poilblanc, T. Ziman, J. Bellissard, F. Mila, and G. Montambaux, Poisson vs. GOE
                Statistics in Integrable and Non-Integrable Quantum Hamiltonians, Europhys. Lett.,
                {\bf 22 (7)}, 537 (1993).
        \bibitem{gaspard} P. van Ede van der Pals and P. Gaspard, Two-dimensional quantum spin Hamiltonians:
                Spectral properties, Phys. Rev. E {\bf 49}, 79 (1994).
        \bibitem{georgeot} B. Georgeot and D.L. Shepelyansky, Integrability and Quantum Chaos in Spin Glass Shards,
                Phys. Rev. Lett. {\bf 81}, 5129 (1998).
        \bibitem{kudo} K. Kudo and T. Deguchi, Level statistics of XXZ spin chains with a random magnetic field,
                Phys. Rev. B {\bf 69}, 132404 (2004).
        \bibitem{deguchi} T. Deguchi, P. K. Ghosh, and K. Kudo, Level statistics of a pseudo-Hermitian Dicke model,
                Phys. Rev. E {\bf 80}, 026213 (2009).
        \bibitem{santos} L. F. Santos and M. Rigol, Onset of quantum chaos in one-dimensional bosonic and fermionic systems
                and its relation to thermalization, Phys. Rev. E {\bf 81}, 036206 (2010).
        \bibitem{gubin} A. Gubin and L. F. Santos, Quantum chaos: An introduction via chains of interacting
                spins 1/2, Am. J. Phys. {\bf 80}, 246 (2012).
	\bibitem{oganesyan} V. Oganesyan and D. A. Huse, Localization of interacting fermions at high temperature,
                Phys. Rev. B {\bf 75}, 155111 (2007).	
\bibitem{gui} L. Qiong-Gui, Anisotropic Harmonic Oscillator in a Static Electromagnetic Field,
Commun. Theor. Phys. {\bf 38}, 667 (2002).
\bibitem{rebane} T. K. Rebane, Two Dimensional Oscillator in a Magnetic Field,
Journal of Experimental and Theoretical Physics, {\bf 114}, 220(2012).

\bibitem{susy} Pijush K. Ghosh, Supersymmetric quantum mechanics on noncommutative space,
Eur. Phys. J. C {\bf 42}, 355(2005).
\bibitem{stopo} D. McDuff and D. Salamon, Introduction to Symplectic Topology, Oxford Science Publications
(1998).
\end{thebibliography}
\end{document}